\documentclass[11pt]{article}
\pdfoutput=1
\usepackage{amsmath,amssymb,epsfig}
\usepackage{graphicx}
\usepackage{xcolor}
\usepackage[utf8]{inputenc}

\newcommand{\be}{\begin{equation}}
\newcommand{\ee}{\end{equation}}
\newcommand{\ea}{\end{array}}
\newcommand{\beqa}{\begin{eqnarray}}
\newcommand{\eeqa}{\end{eqnarray}}

\def\CP2{{\mathbb C}P^2}

\def\CDalign#1{\bgroup\vcenter\bgroup\tabskip 2pt 
       \baselineskip 14pt \lineskip 3pt \lineskiplimit 3pt
       \halign\bgroup &\hfill$##$\hfill\crcr
       #1\crcr\egroup\egroup\egroup} 

\newcommand{\gapproxeq}{\lower .7ex\hbox{$\;\stackrel{\textstyle
>}{\sim}\;$}}
\newcommand{\lapproxeq}{\lower .7ex\hbox{$\;\stackrel{\textstyle
<}{\sim}\;$}}

\newcounter{appendice}

\def\thebibliography#1{{\bf REFERENCES\markboth
 {REFERENCES}{REFERENCES}}\list
 {[\arabic{enumi}]}{\settowidth\labelwidth{[#1]}\leftmargin\labelwidth
 \advance\leftmargin\labelsep
 \usecounter{enumi}}
 \def\newblock{\hskip .11em plus .33em minus -.07em}
 \sloppy
 \sfcode`\.=1000\relax}

\begin{document}
\begin{titlepage}
\title{A phase space description of the FLRW quantum cosmology \\ in Ho$\check{\rm r}$ava-Lifshitz type gravity
\author{
Rub\'en Cordero$^a$\footnote{Email: cordero@esfm.ipn.mx}\,,
 Hugo Garc\'{\i}a-Compe\'an$^{b,c}$\footnote{Email:
compean@fis.cinvestav.mx}\,\, and Francisco J. Turrubiates$^a$\footnote{Email:
fturrub@esfm.ipn.mx}  \\
{\small\it $^a$Departamento de F\'\i sica, Escuela Superior de F\'\i sica
y Matem\'aticas}\\
{\small\it del Instituto Polit\'ecnico Nacional, Unidad Adolfo L\'opez Mateos,} \\
{\small\it Edificio 9, 07738 Ciudad de M\'exico, M\'exico.}\\
{\small\it $^b$Departamento de F\'{\i}sica, Centro de Investigaci\'on y de Estudios Avanzados del IPN}\\
{\small\it P.O. Box 14-740, 07000 Ciudad de M\'exico, M\'exico.}\\
{\small\it $^c$Departamento de F\'\i sica, Divisi\'on de Ciencias e Ingenier\'{\i}as,}\\
{\small\it Universidad de Guanajuato, Campus Le\'on, Loma del Bosque No. 103,}\\
{\small\it Fraccionamiento Lomas del Campestre, Le\'on, Guanajuato, M\'exico.}}}

\date{}
\maketitle
\thispagestyle{empty}
\begin{abstract}
Quantum cosmology of the Friedmann-Lema\^{i}tre-Robertson-Walker model with cosmological constant in the Ho$\check{\rm r}$ava-Lifshitz type gravity is studied in the phase space by means of the Wigner function. The modification of the usual general relativity description by the Ho$\check{\rm r}$ava-Lifshitz type gravity induces a new scenario for the origin of the Universe with an embryonic era where the Universe can exist classically before the tunneling process takes place and which gives rise to the current evolution of the Universe.
The Wigner functions corresponding to the Hartle-Hawking, Vilenkin and Linde boundary conditions are obtained by means of numerical calculations. In particular three cases were studied for the potential of the Wheeler-DeWitt equation: tunneling barrier with and without embryonic era and when the potential barrier is not present. The quantum behavior of these three cases are analyzed using the Wigner function for the three boundary conditions considered.
\end{abstract}

\end{titlepage}
\section{Introduction}
General relativity has been a very successful classical theory
however the efforts to quantize the theory have found serious
problems. For example, a perturbative loop expansion for gravity
posses ultraviolet divergent Feynman diagrams. One possible solution
to this problem is to fix an infinite number of free parameters in
order to have a well-defined ultraviolet structure but the final
result is a theory which is not adequate to describe gravity at
small distance scales because the theory has no predictive power.
Due to this fact gravity is non-renormalizable. The Ho$\check{\rm
r}$ava-Lifshitz (HL) formulation has as a main goal to get a
renormalizable theory by means of higher spatial-derivative terms of
the curvature which are added to the Einstein-Hilbert action
\cite{Horava:2008ih,Horava:2009uw,
Mukohyama:2010xz,Sotiriou:2010wn,Gumrukcuoglu:2011xg,Lepe:2014fda}.

Although Ho$\check{\rm r}$ava-Lifshitz models improve very well
ultraviolet behavior they possess a violation of Lorentz invariance
at ultra-high momenta. This violation is a consequence of the
anisotropic scaling between space and time \cite{Vakili:2013wc}. Due
to the asymmetry of space and time it is very convenient to write
the spacetime metric in terms of the ADM variables where the theory
is called projectable or non-projectable depending if the lapse
function $N$ is only a function of the time coordinate or space and time
coordinates respectively \cite{Blas:2009qj,Blas:2010hb}.

It is important to remark that the projectable theory has an
additional scalar degree of freedom which turns out to be unstable
in the infrared (IR) limit when the running constant $ \lambda > 1$
or $\lambda < 1/3$ and it is a ghost when $ 1/3 < \lambda < 1$
\cite{Sotiriou:2010wn} (see below Eq. \ref{HLaction}). In order to
fix the IR instability for $\lambda >1$, it has been introduced
higher order derivatives in the model that can cut off these
instabilities. However, it has been shown
\cite{Charmousis:2009tc,Blas:2009yd,Bogdanos:2009uj,Koyama:2009hc}
that the scalar mode is strongly coupled and a perturbative
calculation is not consistent when $ \lambda \to 1$ in the IR limit.
In this situation, the graviton dynamics at very low energies is
modified by the higher order operators resulting in a disagreement
with the current observations.  In Refs.
\cite{Mukohyama:2010xz,Izumi:2011eh,Gumrukcuoglu:2011ef}, this IR
instability was studied perturbatively by using the gradient
expansion method for three different gravitational settings. In there it is found that general relativity (GR)
plus dark matter (DM) is restored in the limit $\lambda \to 1$ by
nonlinear dynamics. In \cite{Mukohyama:2010xz} it was discussed the
gradient expansion about an spherically symmetric static solution
and therefore the DM part does not contribute as it is a
time-dependent solution. Moreover in Ref. \cite{Izumi:2011eh} this
expansion is performed in the context of pure gravity about a
cosmological solution of the Friedmann-Lema\^{i}tre-Robinson-Walker
(FLRW) type. The case including a scalar field coupled to gravity
was described in \cite{Gumrukcuoglu:2011ef}.   A generalization
including the non-projectable case was discussed in this context, in
Ref. \cite{Fukushima:2018xgv}.

In the non-projectable models, additionally to the invariants
present in the action constructed with the spatial metric
${^{(3)}}g_{ij}$, it is possible to consider invariant contractions
of the quantities $a_i = \frac{\partial \ln N}{\partial x^i}$. In
the case of the lower order invariant $a^i a_i$, there exists a
parameter $\sigma$ that characterize an adequate domain of the
theory \cite{Sotiriou:2010wn,Sotiriou:2009bx}. When $0<\sigma<2$ and
$\lambda >1$, there is also an extra scalar degree of freedom which
is classically stable and it is not a ghost. It should be noticed
that the non-projectable model has a strong coupling
\cite{Sotiriou:2010wn,Charmousis:2009tc,Papazoglou:2009fj} although
it is not accessible from gravitational experiments
\cite{Sotiriou:2010wn}.

However, in the minisuperspace approximation for
cosmological models where homogeneity and isotropy are the main
ingredients many of the results obtained in the projectable case
also apply to the non-projectable version and vice versa. For
example, in the projectable version of the theory the Hamiltonian
constraint is non-local and in the non-projectable case it is local.
The difference between both theories is reflected in the classical
dynamics of the scale factor in an extra term of the type of non
relativistic matter that could be selected to be zero
\cite{Sotiriou:2010wn}. For this reason in the minisuperspace
cosmology approximation the non-projectable or projectable
Ho$\check{\rm r}$ava-Lifshitz gives the same results in this case.
On the other hand, the problem of the instabilities in the case of
the projectable HL gravity does not appear when we consider a
homogeneous and isotropic spacetime. It is important to recall that
even at the classical level this kind of instabilities could be
emerged when the perturbations are calculated around the homogeneous
and isotropic spacetime.

Another characteristic of the Ho$\check{\rm r}$ava-Lifshitz model is the existence of the detailed
balance condition (which is inspired by condensed matter system) where the potential term in the action is constructed with the aid of the variation of a superpotential with respect to the spatial metric. In the three dimensional case the potential term can be written in a special combination of
covariant derivatives of the Ricci tensor and the scalar curvature
which has as a special feature that the theory have mild
renormalizable properties \cite{Vakili:2013wc}. Nevertheless,
despite the detailed balance system has an easier quantum
characteristics (since the number of terms that could be considered are reduced with the introduction of the superpotential), the non-detailed balance model with additional terms
in the Lagrangian has the nicer behavior which allows to recover
the detailed balance model. In addition, the non-detailed balance model is phenomenologically better behaved than the model with the detailed balance condition since in the classical limit the cosmological constant and Newton constant are independent and they could be adjusted to confront with observations. Taking into
account these characteristics it is interesting to study models
without the detailed balance condition.

Considering that Ho$\check{\rm r}$ava-Lifshitz model is a new theory of gravity, it turned out
very important to investigate its consequences at the cosmological level. Several aspects
of Ho$\check{\rm r}$ava-Lifshitz cosmology have been studied for example
in \cite{Wang:2009rw,Yamamoto:2009tf,Maeda:2009hy,Carloni:2009jc,Wang:2009azb,Gao:2009wn,Dutta:2010jh,Saridakis:2011pk}.
Moreover, since Ho$\check{\rm r}$ava-Lifshitz gravity produces modifications of general relativity in the ultraviolet limit, it is very relevant to study its implications in the process of the birth of the Universe.
Quantum cosmology has tried to understand the
mechanism that gives rise to the origin of the Universe by using quantum
properties during the first stages of the evolution of the Universe.
The first steps in the development of quantum cosmology started at
the beginning of the 80's of last century where it was proposed that
the Universe could be spontaneously nucleate out from nothing
\cite{DeWitt,Tryon,Fomin,Atkatz,Alex1,Zeldovich,HawkingT,LindeT,Rubakov,AlexT}.
The evolution can be described in the following way. After
nucleation, the Universe can enter a phase of inflationary
expansion. At the end of its exponential expansion it continues its
evolution until the present time following, for example, the
description provided by the standard cosmological scheme. It is
important to mention that there are several essential issues that
remain to be solved like the general definition of probability, time
and boundary conditions \cite{approch}. Thus, one of the principal
goals is to find a unique solution of the Wheeler-DeWitt
differential equation as well as to impose boundary
conditions. In ordinary quantum mechanics there is an external
system and the boundary conditions can be enforced safely, but in
4-dimensional quantum cosmology there is nothing external to the
Universe and the issue of which one is the correct choice for the
boundary condition of the Universe had an answer open to debate
\cite{debate}. There are several choices for the right boundary
conditions in quantum cosmology, for example, the no-boundary
proposal of Hartle and Hawking \cite{HawkingT}, the tunneling
proposal of Vilenkin \cite{AlexT} and the proposal of Linde
\cite{LindeT}. Recent results in quantum cosmology
\cite{Brustein:2005yn} were obtained through the principle of
selection in the landscape of string vacua.
Minisuperspace formulation in Ho$\check{\rm r}$ava-Lifshitz quantum
cosmology has been developed in \cite{Bertolami:2011,Pitelli:2012,
Christodoulakis:2011np,Obregon:2012bt,Vakili:2013wc,Benedetti:2014dra}
where classical and quantum solutions are obtained and a detailed analysis of
the dynamics is performed. Besides, in \cite{Vakili:2013wc}
the Wheeler-DeWitt equation is found and the corresponding wave
functions are studied.

On the other hand, an alternative scheme to describe quantum systems
is given by the phase space quantum mechanics. This framework
provides a different approach to the usual treatment performed in
only one representation (coordinates or momenta). In fact, it gives
a different perspective of the quantum phenomena since the relations
between the coordinates and momenta can be analyzed simultaneously. A
complete and detailed review of this construction can be found in
\cite{Kim,CFZ2}. In this approach the central role is played by the Wigner function
in terms of which all the quantum information of the system can be
obtained. This function in principle offers the possibility of study
the semiclassical properties as well as the classical limit in a
more direct way. The use of the Wigner function has had a great
interest in different areas of research like in statistical physics,
nuclear and particle physics, quantum optics, condensed matter and in signal
processing among others (see \cite{Weinbub,Dragoman}). Lately, different techniques have been proposed to
measure this function experimentally \cite{Kurtsiefer,Ourjoumtsev,Deleglise}.

The phase space quantum description has already been employed to
treat certain cosmological models under the Einstein gravity
formulation \cite{Cordero:2011xa}.

Recently an analysis of the FLRW cosmological model in the HL gravity using
the Wigner function was given by Bernardini, Leal and Bertolami \cite{Bernardini:2017jlz}. In their article,
they take in consideration the time evolution and showed that the association of the
quantum variable of time with the radiation energy is a suitable option to move from the
quantum to the classical behavior. This transition is studied by means of the Wigner flows
since they affirm that the classic limit is easier to obtain than from the Wigner
functions.
\\
In this paper we investigate also the quantum properties in the phase space of the FLRW model with
cosmological constant in the HL gravity using the Wigner function. However in our treatment we do not use
asymptotic or perturbative solutions of the Wheeler-DeWitt equation for the HL theory. We study directly the Wigner functions obtained
from the wave functions of the Wheeler-DeWitt equation for the full potential in the HL theory. Moreover we do not employ
the quasi-Gaussian superposition of these states that was used by Bernardini et al. We follow the procedure
that we used in a previous work \cite{Cordero:2011xa} where this cosmological model was treated in the framework
of the usual general relativity. Our aim is to examine the differences obtained for the FLRW model in
the phase space when the effects of the HL gravity are considered for different boundary conditions as well
as the physical consequences of a tunneling barrier that appears for certain values of the HL parameters.
In the HL type gravity the corresponding potential of the Wheeler-DeWitt equation can be parameterized by only
two factors which allows to classify it in four possible behaviors. The different cases for the potential
of the complete Wheeler-DeWitt equation which present a potential barrier are considered and each of them is
studied for the Hartle-Hawking, Vilenkin and Linde boundary conditions. In particular, an embryonic epoch can occur
where the universe could exist classically before the tunneling process that gives rise to the current Universe.

The paper is organized as follows. In Section 2 we
present the main features of the Ho$\check{\rm r}$ava-Lifshitz
gravity and the corresponding Wheeler-DeWitt equation. A brief
description of quantum mechanics in the phase space is presented in
Section 3, mainly the construction of the Wigner function, which is
the principal element that will be used later. In Section 4 we
show the parameter region which gives a tunneling behavior for
the quantum potential. In particular, we find a very interesting
scenario where the Universe could be in a kind of embryonic epoch
where a classical state is possible before tunneling through a potential
barrier and nucleate with a finite value (a similar behavior appears in
brane quantum cosmology \cite{Davidson}). The embryonic era is not
present in usual general relativity. Besides we calculate numerically the wave and
Wigner functions from the solutions of the complete Wheeler-DeWitt
equation and analyze the quantum behavior of the Universe
during this process for three different boundary conditions: the
Hartle-Hawking no boundary proposal, the tunneling or
Vilenkin boundary condition and the Linde condition. Finally in
Section 5 we give our final remarks.

\section{Ho$\check{\rm r}$ava-Lifshitz gravity}
\setcounter{equation}{0} The Ho$\check{\rm r}$ava-Lifshitz
formulation of gravity is an alternative theory to general
relativity which employs higher spatial-derivative terms of the
curvature which are added to the Einstein-Hilbert action with the
aim of obtaining a renormalizable theory. We consider the action for the projectable model which
is given by \cite{Vakili:2013wc,Bertolami:2011,Pitelli:2012,Bernardini:2017jlz,Maeda:2010ke}
\begin{eqnarray}
S &=& \frac{M_{Pl}^2}{2} \int_M d^3x dt N \sqrt{h} \bigg[K_{ij}K^{ij}
- \lambda K^2 -g_1 R - g_0 M_{Pl} ^2 -M_{Pl}^{-2} \big(g_2 R^2 + g_3 R_{ij}
R^{ij}\big) \nonumber \\
&-& M_{Pl}^{-4} \big( g_4 R^3 + g_5 R R^i_{ \ j}R^j_{ \ i} + g_6 R^i_{
\ j}R^j_{ \ k}R^k_{ \ i} + g_7 R \nabla^2R + g_8 \nabla_i R_{jk}
\cdot \nabla^i R^{jk} + g_9 \epsilon^{ijk}R_{il} \nabla_j R^{l}_{k} \big)\bigg]
\nonumber \\
&+& M_{Pl}^2 \int_{\partial M} d^3x
\sqrt{h}K,
\label{HLaction}
\end{eqnarray}
where $g_i$ $(i= 0,1,...,9)$ are dimensionless running couplings constants,
$M_{Pl}=(8\pi G)^{-1/2}$ is the Planck mass, $K$ is the trace of
$K_{ij}$ which are the components of the extrinsic curvature, $h$ is
the determinant of the spatial metric $h_{ij}$, $R$ is the scalar
curvature, $R_{ij}$ the Ricci tensor of the spatial geometry,
$\varepsilon^{ijk}$ is the Levi-Civita symbol and $N$ denotes the
lapse function. The running constants $g_i$ and $\lambda$ give the HL
corrections to general relativity.  The general relativity action is
in principle recovered (in the limit when the curvature radius is much
bigger than the Planck length) if we define $g_0= 2\Lambda M_{Pl}
^{-2}$, $g_1= -1$, $\lambda =1$ and $g_i=0$ for $i \geq 2$. It is
important to mention that in this formulation $\lambda$ should be
considered as a running constant which represents the infrared limit
of the gravitational theory.

Actually the GR limit is quite subtle. Here we will briefly
comment about this subtle limit for the projectable theory. In HL
gravity the complete action includes higher-dimensional
operators, given by the higher-order curvature terms. The
perturbative analysis (around a given classical solution) of the
whole system leads to find that the scalar degree of freedom does
not decouple in the IR regime to obtain GR. In our case, as it was mentioned in the Introduction, we would
have to make a similar analysis as the one described in Ref.
\cite{Izumi:2011eh}. Thus, in this case, a possible discontinuity
at $\lambda =1$ does not appear due the same arguments of
\cite{Izumi:2011eh}, i.e. the limit is restored by nonlinear
dynamics. The expansion about a time-dependent solution of the FLRW
type prevents us to neglect the DM effect at the level of the
Friedmann equation in the whole gravitational dynamics of the HL
theory. However in our paper we are considering the dynamics given
at the level of the Hamiltonian constraint which is given by the
Wheeler-DeWitt equation. Therefore at this level we are looking for
to solve the constraint equation and it is still not necessary to
introduce the Friedmann equation and to take into account the DM
effect. Thus from this point of view the projectable theory give us
results that does not take into account the scalar mode. \\
Finally, in Ref. \cite{Barvinsky:2015kil} the renormalizability of
the projectable HL gravity was performed. It was argued that due the
local gauge fixing in this theory, there will arise some
modifications of the propagators of the metric. However it was shown
there that this effect does not spoil the convergence of the loop
integrals. Even that the projectable theory does not reproduce GR in
the IR limit, the mentioned work \cite{Barvinsky:2015kil} is
valuable as a preliminary understanding if one is looking for to
carry out the renormalizability in the non-projectable case, which
certainly reproduces GR.

We are interested in the FLRW cosmology in the context of
Ho$\check{\rm r}$ava-Lifshitz theory of gravity and we consider the
usual FLRW metric given by
\begin{equation}
 ds^2 = -N^2 dt^2 + a^2(t) \left[ \frac{dr^2}{1-kr^2}+ r^2(d\theta^2+ \sin^2\theta d\varphi^2)\right],
\end{equation}
where $a(t)$ denotes the scale factor, $k$ the curvature of the
spatial section and $r,\theta,\varphi$ the 3 dimensional spherical
coordinates. In order to calculate the action for the FLRW metric it
is necessary to compute the extrinsic curvature
\be K_{ij} =
-\frac{1}{2N} \frac{ \partial h_{ij}}{\partial t},
\ee
since the shift vector is zero. For the spatial geometry the following results
are very useful $K_{ij} K^{ij} = \frac{3 \dot{a} ^2}{N^2 a^2}$, $K=
-\frac{3 \dot{a} }{N a}$, where the dot stands for the derivative
with respect to time. The Ricci tensor and the Ricci scalar are
given respectively by $\frac{2k h_{ij}}{a^2}$ and $\frac{6k}{a^2}$.
The gravitational part for FLRW is given by
\begin{eqnarray}
 S_g &=& \frac{3V_0 M_{Pl}^2 (3\lambda -1)}{2}\int dt N \left\{ -\frac{a\dot{a}^2}{N^2}+ \frac{6ka}{3(3\lambda -1)} - \frac{2\Lambda a^3}{3(3\lambda -1)} \right.\\ \nonumber
 &-& \left. M_{Pl}^{-2} \left[\frac{12k^2(3g_2 +g_3)}{3a(3\lambda -1)}\right]-M_{Pl}^{-4} \left[\frac{24k(9g_4+3g_5 + g_6)}{3a^3(3\lambda -1)}\right]\right\},
\end{eqnarray}
where $V_0 = \int d^3x \sqrt{h}$. Making the selection $\frac{3V_0 M^2 _{Pl} (3\lambda -1)}{2} =1$, the effective Lagrangian can be written in the following way
\be
{\cal L} = N\left( -\frac{a {\dot a}^2}{N^2} + g_c ka - g_\Lambda a^3 - \frac{g_r k^2}{a} - \frac{g_s k}{a^3} \right),
\ee
where the coefficients have the values $g_c =\frac{2}{3\lambda -1}$, $g_\Lambda = \frac{2\Lambda}{3(3\lambda -1)}$, $g_r = 6V_0(3g_2 + g_3)$ and $g_s = 18V_0 ^2(3\lambda -1)(9g_4 + 3g_5 + g_6)$. The terms that include the coefficients $g_7,g_8$ and $g_9$ identically vanish since the scalar curvature only depends on time and the Ricci tensor is proportional to the spatial metric.
The Hamiltonian for the gravitational part then can be calculated by means of the Legendre transformation $H= \dot{a} P_a - {\cal L}$, where the canonical momentum is
\be
P_a = \frac{\partial \cal{L}}{\partial \dot{a}} = -2 \frac{a{\dot a}}{N}.
\label{momentum}
\ee
In this way the Hamiltonian has the following form
\begin{equation}
H = N\left[ -\frac{{P_a}^2}{4a} -g_cka + g_\Lambda a^3 + \frac{g_r k^2}{a} + \frac{g_s k}{a^3} \right] .
\label{localhamiltonian}
\end{equation}
Now, some remarks are appropriate at this part. As was mentioned before, in the projectable version of the theory the Hamiltonian constraint is non-local and in the non-projectable case it is local. It is possible to convert the non-local Hamiltonian constraint to a local one when the spacetime is isotropic and homogeneous \cite{Sotiriou:2009gy}. In fact, the local Hamiltonian constraint (\ref{localhamiltonian}) is the same that was used in other previous papers which investigate quantum cosmology in HL gravity \cite{Vakili:2013wc,Bertolami:2011,Pitelli:2012,Bernardini:2017jlz}. Then, our results should be hold for the non-projectable version in the case when the spacetime is isotropic and homogeneous.
The quantum description of the Universe could be done through the canonical quantization method which gives the Wheeler-DeWitt equation as follows \cite{Vakili:2013wc}
\begin{equation}
\left[\frac{1}{a} \frac{d^2}{da^2} - \frac{p}{a^2}\frac{d}{da}+ 4\left( -g_{c}ka + g_\Lambda a^3 + \frac{g_r k^2}{a}+ \frac{g_s k}{a^3}\right)\right] \Psi (a) =0 \,\,\,,
\label{WKBC}
\end{equation}
where $\Psi (a)$ is the wave function of the Universe and $p$
represents the ambiguity in the factor ordering of the operators $a$
and $P_a$. With the selection of $p=1$ and performing the change of
variable $q=a^{2}$ the Wheeler-DeWitt equation (\ref{WKBC})
transforms to
\begin{equation}
\left[-\frac{d^2}{dq^2} + V(q) \right] \Psi (q)= 0, \label{WKBCT}
\end{equation}
where
\begin{equation}
V(q)=g_{c}k - g_\Lambda q - \frac{g_r k^2}{q} - \frac{g_s k}{q^{2}},
\label{WDWPOTENTIAL}
\end{equation}
is the Wheeler-DeWitt quantum potential.

It is difficult to obtain exact analytic solutions of the
Wheeler-DeWitt equation (\ref{WKBCT}). Thus in Section 4 we will find
numerical solutions to this complete equation for different values
of the parameters, without any particular or asymptotic case but including large $q$
boundary conditions that we discuss below. These wave functions are then
used to obtain the corresponding Wigner functions which are the main aim
of the present article.

As we know there exist the possibility to find the wave function for
certain limits. For example, for very small values of the scale
factor the first two terms of the potential in the WDW equation are
negligible with respect to the last two terms and the solutions can
be written as \be \Psi(a) = a\left[ A J_{\sqrt{1-4kg_s}}(2k
\sqrt{g_r}a) + B Y_{\sqrt{1-4kg_s}}(2k \sqrt{g_r}a) \right], \ee
where $J_\nu(a)$ and $Y_\nu(a)$ are the Bessel functions and $A$,
$B$ are complex constants.

On the other hand, for large values of the scale factor the first
two terms of the potential are the important ones and the last two
are negligible. In this limit the solutions correspond to the ones
of the general relativity and they are written in terms of a
combination of the Airy functions in the following way
\begin{equation}
\Psi(a) = C \mbox{Ai} \left(\frac{kg_c - g_\Lambda a^2}{g_\Lambda
^{2/3}} \right) + D \mbox{Bi} \left(\frac{kg_c - g_\Lambda
a^2}{g_\Lambda ^{2/3}} \right),
\end{equation}
where $\mbox{Ai}(x)$ and $\mbox{Bi}(x)$ denote the two linear
independent solutions of the Airy equation and $C$, $D$ are two constant
complex numbers.

Now, depending of the boundary conditions that have been chosen, we
can have the Hartle-Hawking wave function
\begin{equation}
\Psi_{HH}(q)= \mbox{Ai}\left( \frac{kg_c - g_\Lambda q}{g_\Lambda
^{2/3}}  \right)\, , \label{HHfunction}
\end{equation}
the Linde wave function
\begin{equation}
\Psi_{L}(q)= -i\mbox{Bi}\left( \frac{kg_c - g_\Lambda q}{g_\Lambda
^{2/3}} \right)\, , \label{Lfunction}
\end{equation}
or the Vilenkin wave function \cite{approch}
\begin{equation}
\Psi_{V} (q) = \frac{1}{2}\mbox{Ai}\left( \frac{kg_c - g_\Lambda
q}{g_\Lambda ^{2/3}} \right) + \frac{i}{2}\mbox{Bi}\left(
 \frac{kg_c - g_\Lambda q}{g_\Lambda ^{2/3}} \right),
\label{Vfunction}
\end{equation}
where the change of variable $q=a^2$ has been employed. One of the
main differences between the Hartle-Hawking, Linde and Vilenkin wave
functions is the behavior for the region $q>kg_s/g_\Lambda$. In such case,
the Hartle-Hawking wave function has the following expression
\begin{equation}
\Psi_{HH} = \psi_{+}(q) + \psi_{-}(q),
\end{equation}
the Linde wave function is
\begin{equation}
\Psi_{L} = \psi_{+}(q) - \psi_{-}(q),
\end{equation}
and the Vilenkin tunneling wave function is written like
\begin{equation}
 \Psi_{V} = \psi_{-}(q),
\end{equation}
where $\psi_{-}(q)$ and $\psi_{+}(q)$ describe an expanding and contracting
Universe, respectively (see \cite{Cordero:2011xa}).

It is important to notice the following. Near the origin the asymptotic behaviour of the wave function can be written in terms of the Bessel functions which diverges or is zero when $a=0$ depending if $A=0$ or $B=0$ respectively. Then, we can select a vanish value for the wave function at the origin, however we are interested to impose boundary conditions at large values of the scale factor. We tried to fulfill both boundary conditions, at the origin with a vanish wave function and some of the former boundary conditions at large values of the scale factor. We explored numerically a set of values for the wave function and its derivatives at large values which satisfy each one of the three boundary conditions and we looked for a wave function that vanish at the origin at the same time, however it was not possible to fulfill these two conditions simultaneously.

\section{Quantum phase space description}
\setcounter{equation}{0}

The phase space quantum description is an alternative approach to the standard construction of quantum mechanics in a Hilbert
space. In this scheme coordinates and momenta are employed together, providing a direct extension of the Hamiltonian construction to
describe quantum systems. Under this framework the information of the system is obtained by the real valued Wigner function which plays an analogous
role as the wave function.  However, since this function can take negative values, it defines a quasi-probability distribution
function as opposed to the squared modulus of the wave function $|\psi|^{2}$ that gives a usual probability distribution. The Wigner
function has encoded all the quantum information of the system and in principle allows the study of its semi-classical
properties as well as the analysis of the classical limit in a more direct way. A detailed review of this topic
can be consulted in Refs. \cite{Kim,CFZ2} and the references cited therein.

It must be pointed out that quantum mechanics in phase space is just part of a more consistent and broader
type of quantization known as \textit{deformation quantization} which it has been an important subject of study in mathematical
physics since its introduction in full form by Bayen et al in 1978 \cite{Bayen}. Under this viewpoint quantization is
understood as a deformation of the usual product algebra of the smooth functions on the classical phase space, which in
turn it induces a deformation of the Poisson bracket algebra. The deformed product is called the \textit{$*$ - product} (star product)
and its existence has been proven first for the case of any symplectic manifold and later for any Poisson manifold \cite{Fedosov:1994zz,
Kontsevich:1997vb}. These results in principle allow us to carry out the quantization of arbitrary Poissonian or symplectic systems
which gives an advantage over the other quantization methods developed until now. An updated and detailed review of this
quantization formalism can be consulted in Ref. \cite{DitoStern}.

So, if we consider a quantum system with just one degree of freedom specified by the density operator
\begin{equation}\label{densityoperator}
 \hat{\rho}=\sum_{j}a_j| \psi_{j}\rangle \langle \psi_{j}|,
\end{equation}
where $a_j$ denote a set of non-negative quantities whose sum is equal
to one, then the corresponding Wigner function can be constructed by its Fourier
transform as follows
\begin{equation}\label{densoperfouriertransf}
  \rho_{W}(x,p)=\frac{1}{2\pi\hbar}\int \big\langle x + \frac{y}{2}|\hat{\rho} |x - \frac{y}{2} \big\rangle \exp \bigg\{\frac{-ipy}{\hbar}\bigg\}  dy.
\end{equation}
In particular for a pure state the density operator has the following form $\hat{\rho}=| \psi\rangle \langle \psi|$, then using the previous
equation the Wigner function can be written by means of the wave function $\psi(x)$ as
\begin{equation}
\rho_{W}(q, p) = \int^{\infty} _{-\infty} \frac{d\xi}{2\pi \hslash} \exp \bigg\{-i \frac{\xi}{\hslash}p \bigg\} \Psi^*\left( q- \frac{\xi}{2}\right) \Psi \left( q + \frac{\xi}{2}\right).
\label{integralexp}
\end{equation}
The former expressions for the Wigner function can be extended
directly to the $\mathbb{R}^{2n}$ phase space.

In a similar way as with the wave function in the usual quantum
mechanics formalism the Wigner function allows to determine the
behavior of the system, and fulfills the following properties:
\begin{itemize}
  \item $\rho_{W}(x,p)=\rho_{W}^{*}(x,p)$, it is a real function.
  \item $\int_{\mathbb{R}^{2}} \rho_{W}(x,p)dxdp=1$, it is normalized.
  \item $\int_{\mathbb{R}} \rho_{W}(x,p)dp= \alpha(x)$, it defines a positive space probability density.
  \item $\int_{\mathbb{R}} \rho_{W}(x,p)dx= \beta(p)$, it gives a positive momentum probability density.
\end{itemize}
In this way, the expectation value of an observable $f=f(x,p)$ under this picture is then given by
\begin{equation}\label{expectvalue}
  \langle f \rangle = \int_{\mathbb{R}^{2}} f(x,p) \rho_{W}(x,p)dxdp.
\end{equation}
In spite of providing an important tool to deal with more complex systems the use of the
Wigner function in quantum cosmology has not been employed widely and our objective is to
explore the properties that it offers and to complete the quantum description given by the
other methods. In the next part of this work we will deal with the model described in
Section 2 by means of the Wigner function.

\section{Wigner function of the FLRW cosmology in HL gravity}

Due to the complexity of the potential (\ref{WDWPOTENTIAL}) it is not possible to obtain analytic solutions
of the complete Wheeler-DeWitt equation (\ref{WKBCT}) so instead its behavior is studied in this section
by means of numerical solutions.

The behavior of this system without cosmological constant and which is also adequate for small values of the scale factor was considered by
Bernardini et al (see \cite{Bernardini:2017jlz}) obtaining analytical solutions of the wave function as well as
for the Wigner function of a quasi-Gaussian superposition of those states. In addition, they performed a perturbative analysis when the term with cosmological
constant is considered. However, we will not make any approximations and we will consider all the
terms of the potential in our treatment.

\begin{figure}
\begin{center}
\includegraphics[scale=1.2]{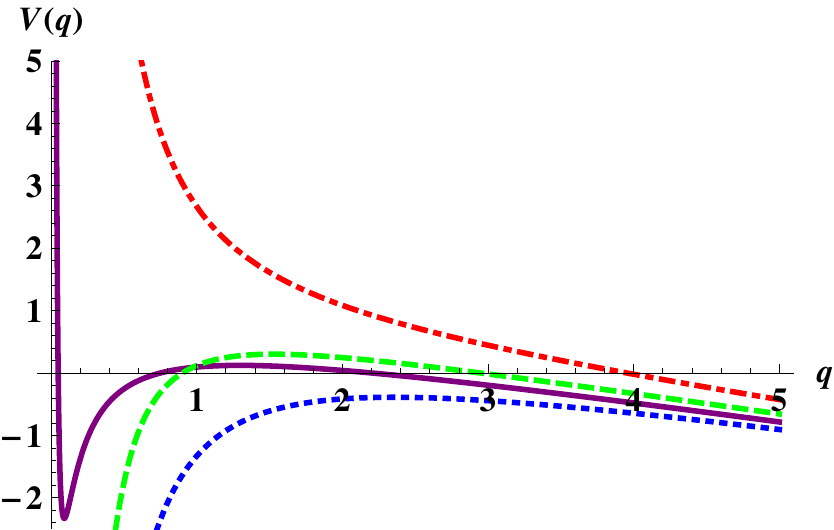}
\caption[Short caption for figure 1]{\label{fig:potential} {\scriptsize The different behaviors of the potential for the Wheeler-DeWitt equation. The purple curve corresponds to the scenario where the Universe presents an embryonic epoch. The dashed green curve represents the case where there is a potential barrier and there is not an embryonic epoch. The dotted-dashed red curve corresponds to the situation where there is not a potential barrier and there is a big bounce. For the case with the dotted blue curve the initial singularity is present and there is not a potential barrier.}}
\end{center}
\end{figure}

It is important to observe first that for different values of the parameters $g_c, g_\Lambda, g_r$ and $g_s$
the potential (\ref{WDWPOTENTIAL}) will present quite different behaviors. Even more the values of the
parameters $g_c, g_\Lambda, g_r$ and $g_s$ can be restricted in fact
to only two effective parameters $\alpha= \frac{g_{r}}{g_{c}}$ and
$\beta= \frac{g_{s}}{g_{c}}$ which give us four different forms of
the potentials showed in Fig. \ref{fig:potential}. For instance, it can present a
tunneling barrier or not as it is shown in Fig. \ref{fig:potential}.
In particular, for certain values of the parameters $\alpha$ and $\beta$, it is possible to
have a potential where the Universe can exists classically when it is very small. This type of
situation is called the {\it embryonic epoch} of the Universe
\cite{Davidson}, and by means of a tunneling process the Universe can
appear in a finite size and expand. This scenario is not present
in general relativity and it is due to the HL quantum corrections terms.

\begin{figure}
\begin{center}
\includegraphics[scale=0.32]{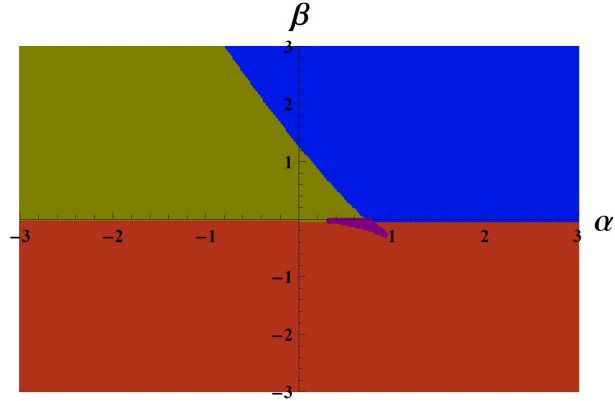}
\caption[Short caption for figure 2]{\label{fig:region} {\scriptsize The plot represents the regions corresponding to different behaviors of the potential for the Wheeler-DeWitt equation (\ref{WKBCT}), where $\alpha= \frac{g_{r}}{g_{c}}$ , $\beta= \frac{g_{s}}{g_{c}}$ and the values $\Lambda=k=1$ were chosen. The purple region corresponds to the scenario where the embryonic epoch is present. The green region represents the situation with a potential barrier without an embryonic era. The red region corresponds to the case without a potential barrier and with a big bounce. The blue sector indicates the scenario where the initial singularity is present and there is not a potential barrier.}}
\end{center}
\end{figure}

The space of parameters $\alpha$ and $\beta$ (with $\Lambda=k=1$) has four well-determined regions as can
be appreciated in Fig. \ref{fig:region}. Each region corresponds to the following cases:
$(a)$ The purple region gives rise to potentials that presents an infinite barrier near the big bang singularity
and an embryonic epoch. $(b)$ The green region produce potentials with a tunneling
structure but without and embryonic era.  Case $(c)$ corresponds to the
blue region which consists of potentials that cannot be zero and without a potential barrier
in which the singularity is present. Finally, case $(d)$ corresponding to the red region presents a big
bounce behavior and consists of potentials that have only one zero.

The former four cases are the only possible behaviors because they are related with the roots of the Wheeler-DeWitt quantum potential (\ref{WDWPOTENTIAL}) which is equivalent to resolve a cubic equation in the $q$ variable. The purple region corresponds to the case of three positive roots for $q$. The green region is associated with two positive roots of the quantum potential. The case of one and zero positive roots correspond to the red and blue regions respectively.

In the next part we will focus on analyzing the cases $(a)$, $(b)$ and $(c)$ since the case $(d)$ does not present
a tunneling process for the Universe and the potential goes to positive infinity when the scale factor tends to zero.

Before we discuss these three cases in the context of the Wigner function we describe the
main features of their associated wave functions. In the process of finding numerical solutions to the
complete Wheeler-DeWitt equation (\ref{WKBCT}) we used the Runge-Kutta 4th order
method and imposed boundary conditions for large values of the scale factor.
For this asymptotic behavior the wave functions tends to the general relativity case but otherwise
the solutions satisfy the full equation (\ref{WKBCT}) depending
of the values of the selected parameters according to Fig. \ref{fig:region}.

\begin{figure}
\begin{center}
\includegraphics[scale=1.2]{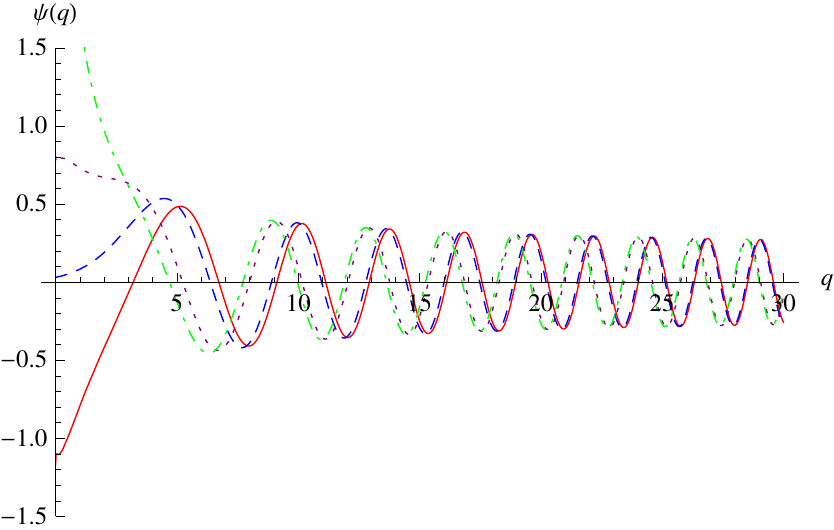}
\caption[Short caption for figure 2]{\label{fig:wftnobigbang} {\scriptsize The wave functions solutions for the case of tunneling with embryonic epoch, purple region case ($\hbar=1$). The red curve corresponds to the numerical solution for the Hartle-Hawking boundary condition considering all the terms included in the Wheeler-DeWitt quantum potential. The dotted purple curve is the numerical solution for the imaginary part of the Linde boundary condition with all the terms in the potential. The dashed blue and dotted-dashed green curves correspond to the $\mbox{Ai}(q)$ and $\mbox{Bi}(q)$ functions respectively. The values of the parameters employed are $g_c$=1, $g_\Lambda$= $\frac{1}{3}$, $g_r$=0.6 and $g_s$=-0.03.}}
\end{center}
\end{figure}

\begin{figure}
\begin{center}
\includegraphics[scale=1.2]{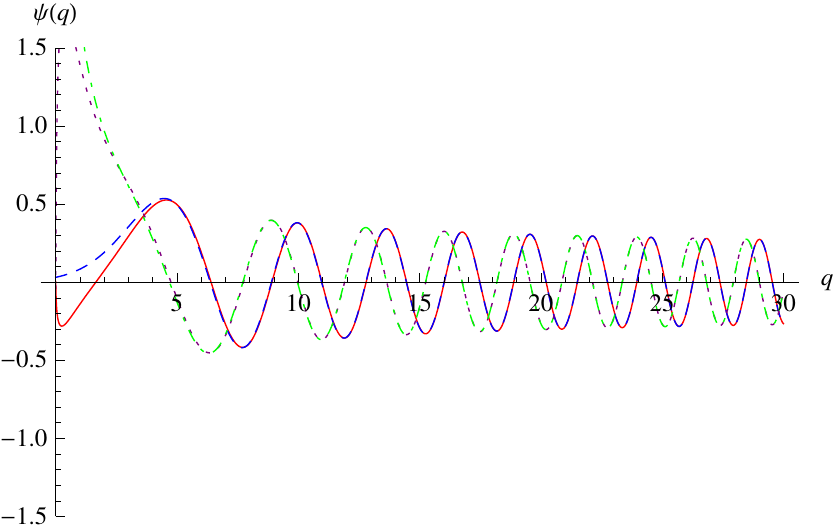}
\caption[Short caption for figure 2]{\label{fig:wftunneling} {\scriptsize The wave functions solutions for the case of tunneling without embryonic epoch, green region case ($\hbar=1$). The red curve represents the numerical solution for the Hartle-Hawking case with all the terms in the Wheeler-DeWitt potential. The dotted purple curve corresponds to the numerical solution for the imaginary part of the Linde boundary condition with all the terms  included in the potential. The dashed blue and dotted-dashed green are the $\mbox{Ai}(q)$ and $\mbox{Bi}(q)$ functions respectively. The values of the parameters employed are $g_c$=1, $g_\Lambda$= $\frac{1}{3}$, $g_r$=0.024 and $g_s$=0.468.}}
\end{center}
\end{figure}

\begin{figure}
\begin{center}
\includegraphics[scale=1.2]{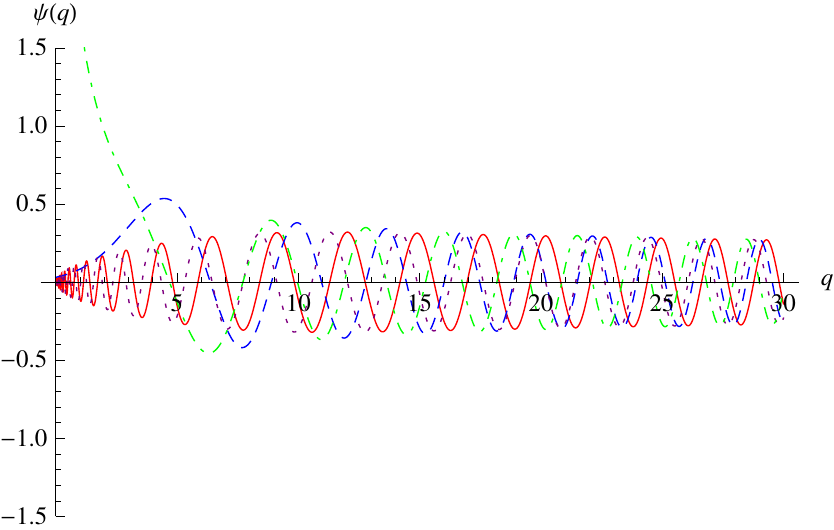}
\caption[Short caption for figure 1]{\label{fig:wfnotunneling}
{\scriptsize The wave functions solutions for the case of no tunneling, blue region case ($\hbar=1$). The red and dotted purple curves represent the numerical solutions for all the terms present in the Wheeler-DeWitt potential for the Hartle-Hawking and the imaginary part of the Linde boundary condition respectively. The dashed blue curve corresponds to the $\mbox{Ai}(q)$ function while the dotted-dashed green curve represents the $\mbox{Bi}(q)$ function. The values of the parameters employed are $g_c$=1, $g_\Lambda$= $\frac{1}{3}$, $g_r$=0 and $g_s$=234.}}
\end{center}
\end{figure}

In Figs. \ref{fig:wftnobigbang}, \ref{fig:wftunneling} and
\ref{fig:wfnotunneling} one can observe the behavior of the wave functions which are
solutions to the equation (\ref{WKBCT}) for two boundary conditions
(Hartle-Hawking and Linde) for a wide range of values of $q$. The wave function for Vilenkin boundary condition
is built, for long distances, from those of Hartle-Hawking and Linde, see equation (\ref{Vfunction}). In these
three figures the curves in red color corresponds to the numerical solutions of
the complete Wheeler-DeWitt equation for the Hartle-Hawking boundary condition
while the dotted purple curves indicates the numerical solutions for the Linde boundary
condition including all the terms of the potential (\ref{WDWPOTENTIAL}).
On the other hand, the curves in dash blue and in dash-dotted green represent respectively the Airy functions
$\mbox{Ai}(q)$ and $\mbox{Bi}(q)$ i.e. the solutions for the asymptotic behavior with large
values of $q$ in the Wheeler-DeWitt potential. As it was mentioned before, we will analyse the Wheeler-DeWitt equation (\ref{WKBCT}) with the complete
potential (\ref{WDWPOTENTIAL}) and we are going to impose the boundary condition
for large values of $q$. The imposition of both boundary conditions at $q=0$ and large values of $q$ at the same time is not
possible in our description.

Figure \ref{fig:wftnobigbang} shows the behavior of wave functions for the case corresponding
to the purple region where there is an embryonic era and tunneling is possible.
In Fig. \ref{fig:wftunneling} we have the case associated to the
green region where the embryonic epoch is not present, there is a potential barrier and
the Universe can arise by tunneling. The Fig. \ref{fig:wfnotunneling} corresponds to
a scenario from the blue region where there is not a barrier and consequently the
Universe cannot start from a tunneling. In all these situations it
can be appreciated that for large values of $q$ the behavior of the
wave functions is the same as in general relativity but for small
values of $q$ the situation is very different. For example, the wave function of
the HL quantum cosmology presents more oscillations near $q=0$
as can be appreciated in Figs. \ref{fig:wftunneling} and \ref{fig:wfnotunneling}. This
is a manifestation that the higher order corrections in the curvature
from the HL theory give rise to a drastic difference with respect
to general relativity.

Now we carry out the analysis of the quantum behavior in the phase space by means of the Wigner function which will be
obtained by a numerical computation employing a Fortran code for the solutions of (\ref{WKBCT}) and using Eq. (\ref{integralexp}).
We will treat separately the $(a)$, $(b)$ and $(c)$ cases mentioned above, each of them for the three boundary conditions considered in section 2.
For simplicity we set $k$ = $V_0$ = $\lambda$ = $\Lambda$ = $g_0$ = $g_1$ =1 which imply that $g_c$ = 1 and $g_\Lambda = \frac{1}{3}$. However, different values of $g_2$,
$g_3$, $g_4$, $g_5$ and $g_6$ were selected for each of the cases that are studied. It is important to note again that such selection was
made primarily for simplicity. The graphics of the Wigner functions along with their corresponding
density plots (where the curves in red denote the classic trajectory) are presented below.


\noindent
$(a)$ {\it Tunneling with embryonic era (Purple Region)}

For this case we employed the following values of the parameters $g_2$ = $g_4$ = $g_5$ = 0, $g_3$ = 0.1 and $g_6$ = -0.0008333 giving the values $g_r$ = 0.6 and $g_s$ = -0.03
which produce values of $\alpha$ and $\beta$ corresponding to a point in the purple region of Fig. \ref{fig:region}, namely a potential barrier with an embryonic epoch.

\begin{figure}
\begin{center}
\includegraphics[scale=0.53]{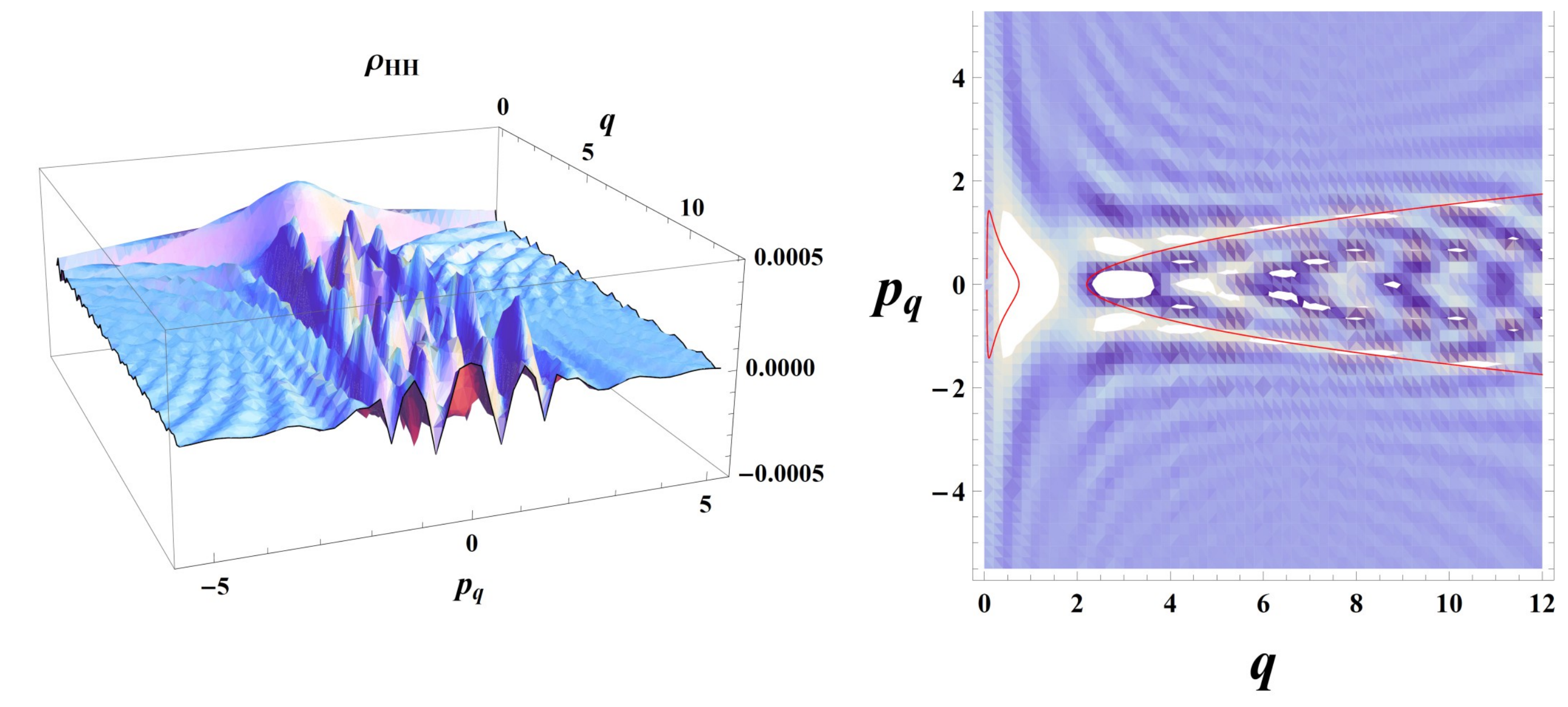}
\caption[Short caption for figure 3]{\label{fig:hhnbb} {\scriptsize The Wigner function and its density plot for the Hartle-Hawking boundary condition with tunneling and embryonic era ($\hbar=1$). The figure shows many oscillations due to the interference between wave functions of expanding and contracting universes. In the density plot it can be observed that the open classical trajectory does not coincide with the highest pick of the Wigner function.}}
\end{center}
\end{figure}

\begin{figure}
\begin{center}
\includegraphics[scale=0.53]{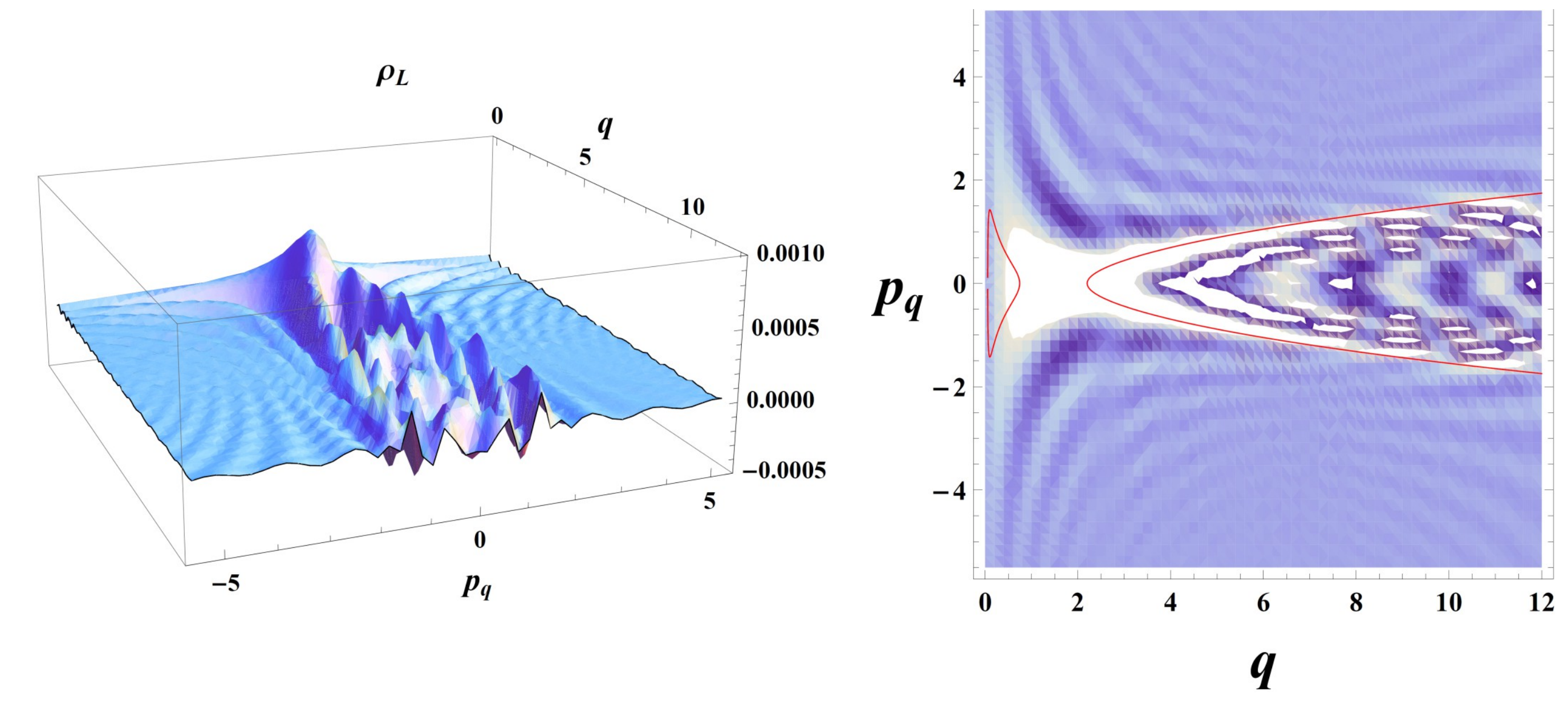}
\caption[Short caption for figure 3]{\label{fig:lnbb} {\scriptsize The Wigner function and its density plot for the Linde boundary condition with tunneling and embryonic era ($\hbar=1$). The figure shows a higher amplitude of the oscillations compared to the Hartle-Hawking case. In the density plot it can be appreciated that the open classical trajectory is near to the higher peaks of his corresponding Wigner function.}}
\end{center}
\end{figure}

\begin{figure}
\begin{center}
\includegraphics[scale=0.53]{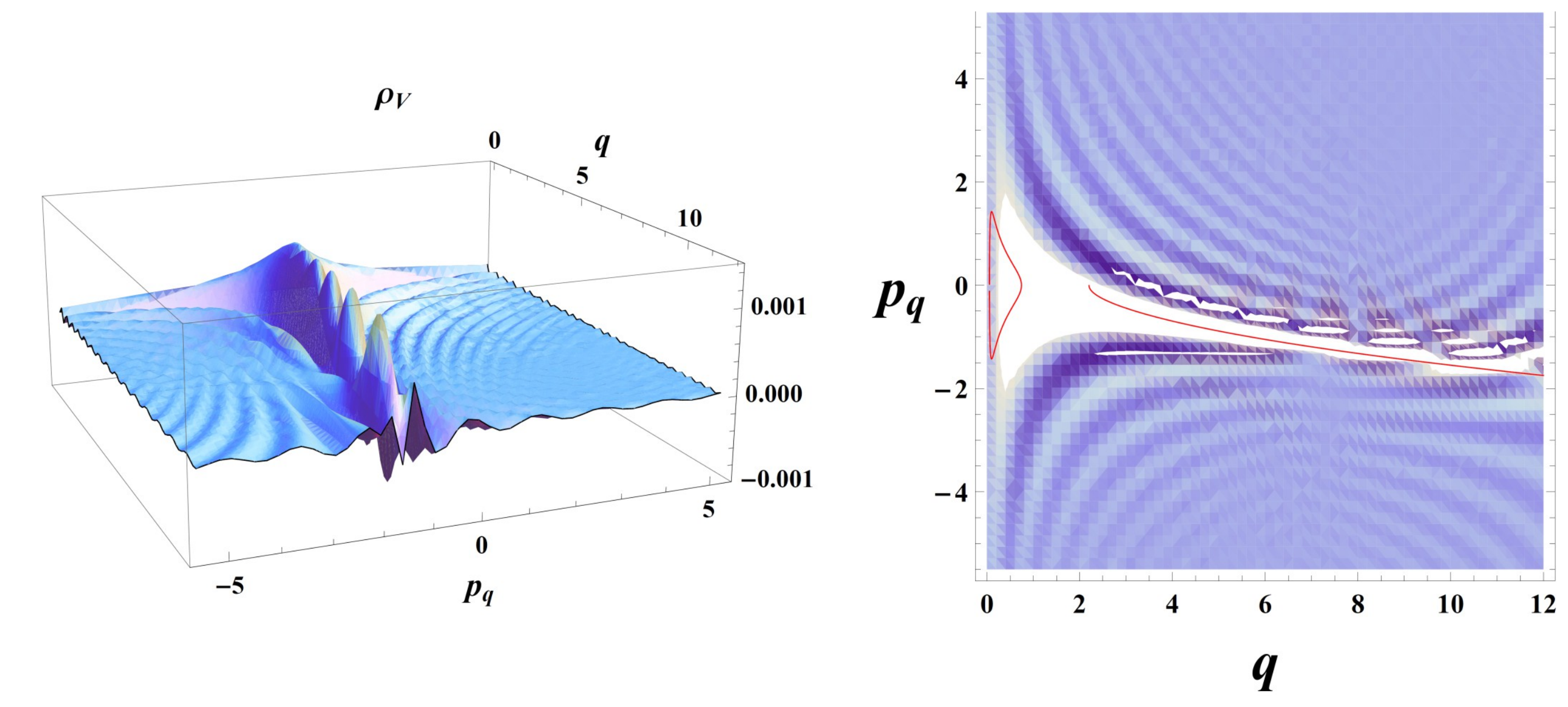}
\caption[Short caption for figure 3]{\label{fig:vnbb} {\scriptsize The Wigner function and its density plot for the Vilenkin boundary condition with tunneling and embryonic era ($\hbar=1$). It can be observed a clear maximum and less oscillations compared with the Hartle-Hawking and Linde cases. The density projection shows that the classical trajectory is at some parts on the maxima of the Wigner function and has only one branch corresponding to an expanding universe.}}
\end{center}
\end{figure}

The results of the Wigner functions as well as their density plots for the Hartle-Hawking wave function of the Eq. (\ref{WKBCT}) are given in Fig. \ref{fig:hhnbb},
for the Linde wave function are presented in Fig. \ref{fig:lnbb} while for the Vilenkin wave function
are shown in Fig. \ref{fig:vnbb}.

The main aspect that can be observed is that for the three boundary
conditions considered the highest peaks of the Wigner functions are
centered around $p_q=0$. Besides, there are two classical trajectories,
one near to the big bang singularity that shows a closed curve which corresponds to the
oscillation between two turning points of the potential describing precisely the embryonic
epoch of the universe while the other corresponds to open trajectories which
represents a contracting (for the upper curve with $P_q >0$ because Eq.(\ref{momentum}))
or expanding (for the lower curve with $P_q <0$) Universe that could arise after a tunneling process.

For the Hartle-Hawking and Linde boundary conditions
it can be appreciated a similar behavior however the Wigner function
has a higher amplitude for the Linde case than for the Hartle-Hawking
boundary condition. Inside the region of the open classical
trajectory more fluctuations are present for the Linde wave function
than for the Hartle-Hawking one, also the highest peaks of the
Wigner function are near the classical trajectories however the
amplitudes decreases with the distance to these trajectories.
Another important difference between these two cases is that the
Hartle-Hawking Wigner function presents more oscillations between the two
classical regions near $p_q=0$ than the Linde Wigner function.
These differences between the two boundary conditions can be
understood taking in consideration that the interference terms have
different sign between a contracting and expanding Universe.

For the Vilenkin boundary condition it can be observed only one
branch which corresponds to an expanding Universe and where the higher oscillations
of the Wigner function are over the open classical trajectory.
This behavior is in agreement with the tunneling boundary condition for
this case. The classical trajectory is in the middle of the higher peaks of the
Wigner function corresponding to negative values of the momenta
which is associated to an expanding Universe. This fact can be
explained in terms of the decoherence of the Vilenkin Wigner
function because there is no interference present between an
expanding and contracting universes like in the other two cases.
A similar analysis for the FLRW model by means of the Wigner function
in the context of general relativity was carried out in
\cite{Cordero:2011xa} but in this work we find an embryonic epoch of the Universe
which constitutes a novel and interesting feature of the HL quantum cosmology.

\noindent
$(b)$ {\it Tunneling without embryonic epoch (Green Region)}

\begin{figure}
\begin{center}
\includegraphics[scale=0.53]{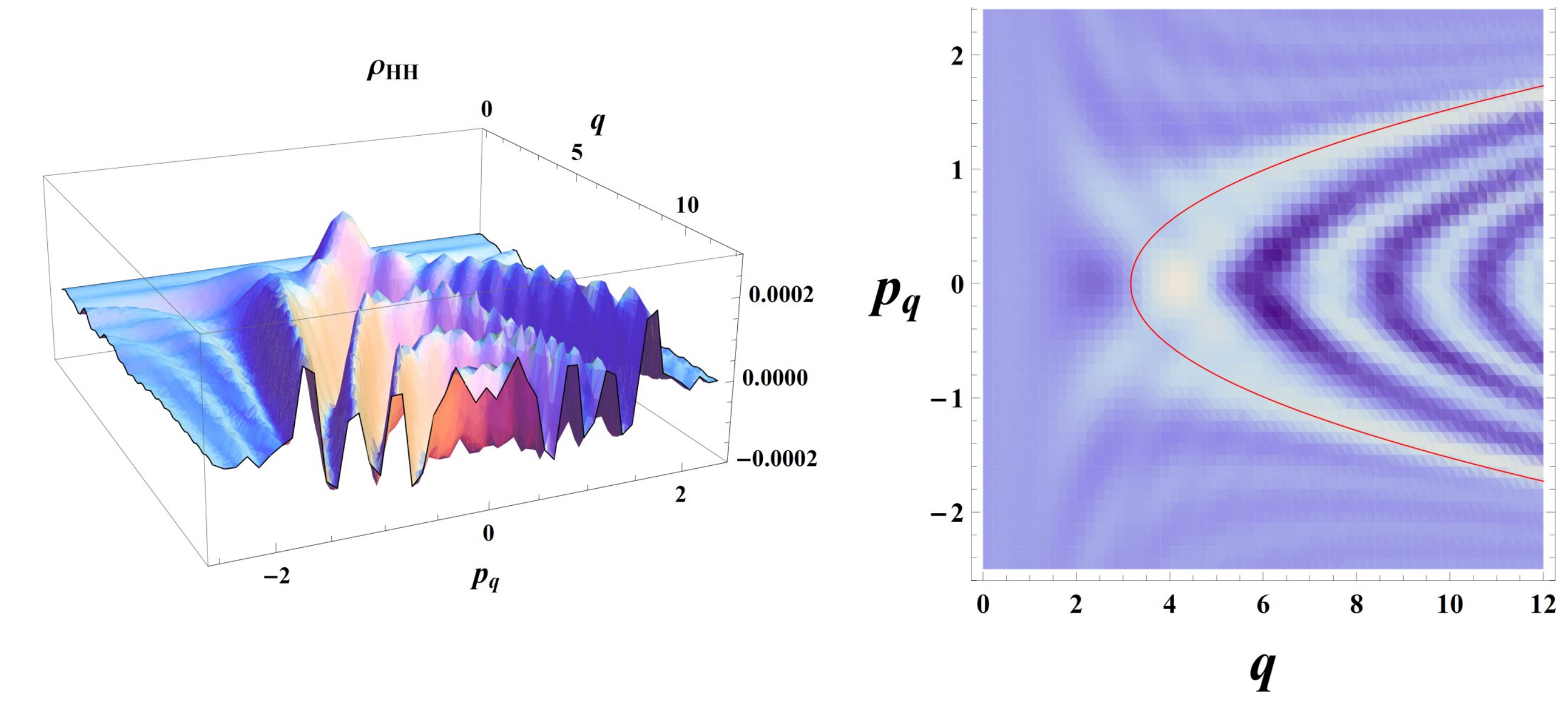}
\caption[Short caption for figure 1]{\label{fig:hh468} {\scriptsize The Wigner function and its density plot for the Hartle-Hawking boundary condition in a scenario where there is a potential barrier without embryonic era ($\hbar=1$). This figure shows many oscillations due to the interference between wave functions of expanding and contracting universes. For the density projection it can be observed that the highest peak is inside of the region bounded by the classical trajectory.}}
\end{center}
\end{figure}

\begin{figure}
\begin{center}
\includegraphics[scale=0.53]{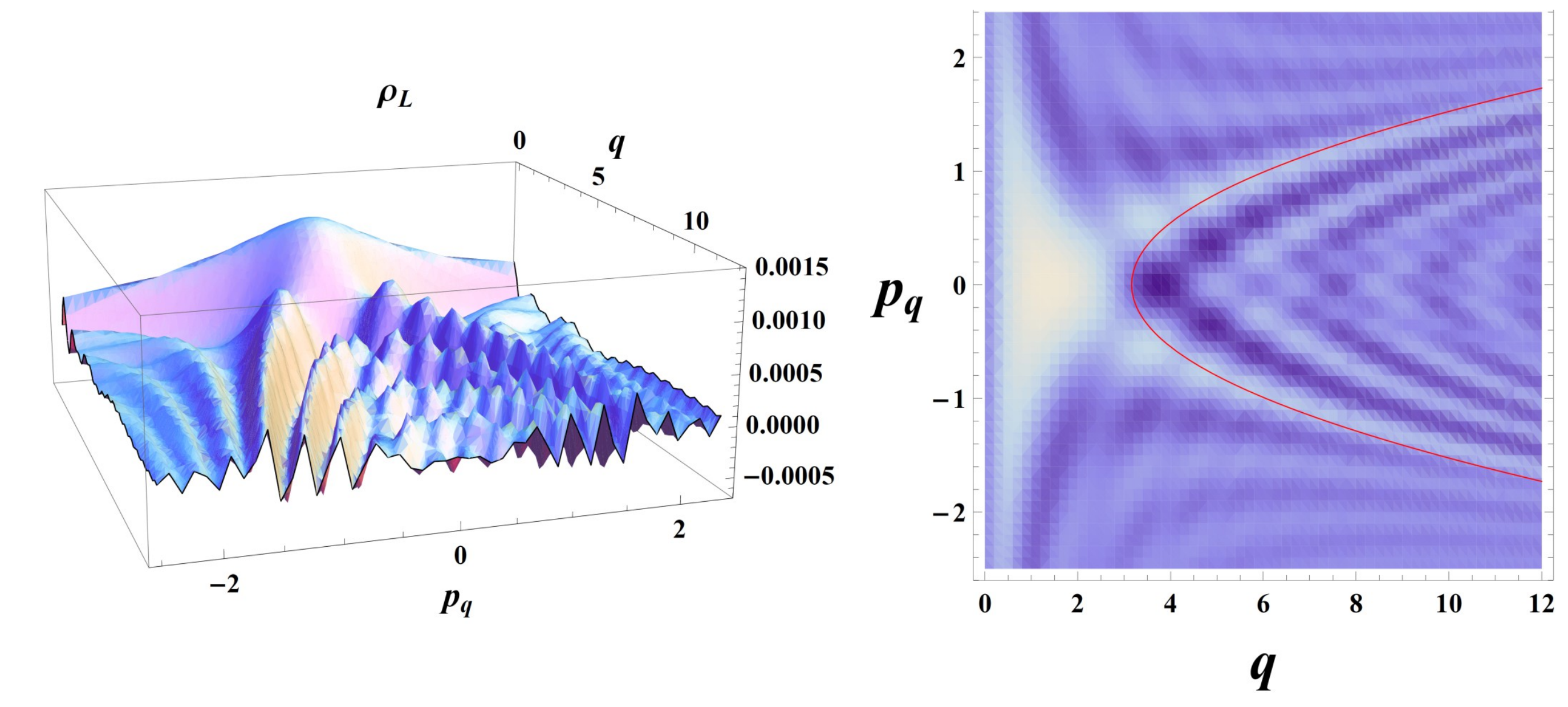}
\caption[Short caption for figure 1]{\label{fig:l468} {\scriptsize The Wigner function and its density plot for the Linde boundary condition in a scenario where there is a potential barrier without embryonic era ($\hbar=1$). The figure shows a higher amplitude oscillations compared to the Hartle-Hawking case. From its density projection it can be appreciated that the highest peak is near the origin of the Universe and outside the region bounded by the classical trajectory.}}
\end{center}
\end{figure}

The second example is given for the next choice of parameters, $g_2$ = $g_3$ = $g_4$ = $g_5$ = $g_6$ = 0.001 so that $g_r$ = 0.024 and $g_s$ = 0.468
which produce a point in the green region of Fig. \ref{fig:region}
and give a typical potential barrier behavior. The corresponding Wigner functions for
the Hartle-Hawking and Linde boundary conditions present similar characteristics like
the classical trajectory is over some of the higher peaks of the Wigner function as well as
that more fluctuations are inside of the region bounded by the classical trajectory.
However, the Linde case posses more oscillations and its amplitudes are higher than in
the Hartle-Hawking case as are shown in Figs. \ref{fig:hh468} and
\ref{fig:l468}. For this case the big-bang singularity is
accessible for the classical and quantum dynamics which is a very important difference
with respect to the FLRW general relativity framework (see \cite{Cordero:2011xa}). In the
Hartle-Hawking case the highest peak is inside the classical region
as it is shown in figure \ref{fig:hh468} in contrast for the Linde
case the highest peak is shifted to the origin of the Universe
outside of the region bounded by the classical trajectory (see
Fig. \ref{fig:l468}). This difference is consequence of the opposite signs in
the interference terms between expanding and contracting universes.

\begin{figure}
\begin{center}
\includegraphics[scale=0.53]{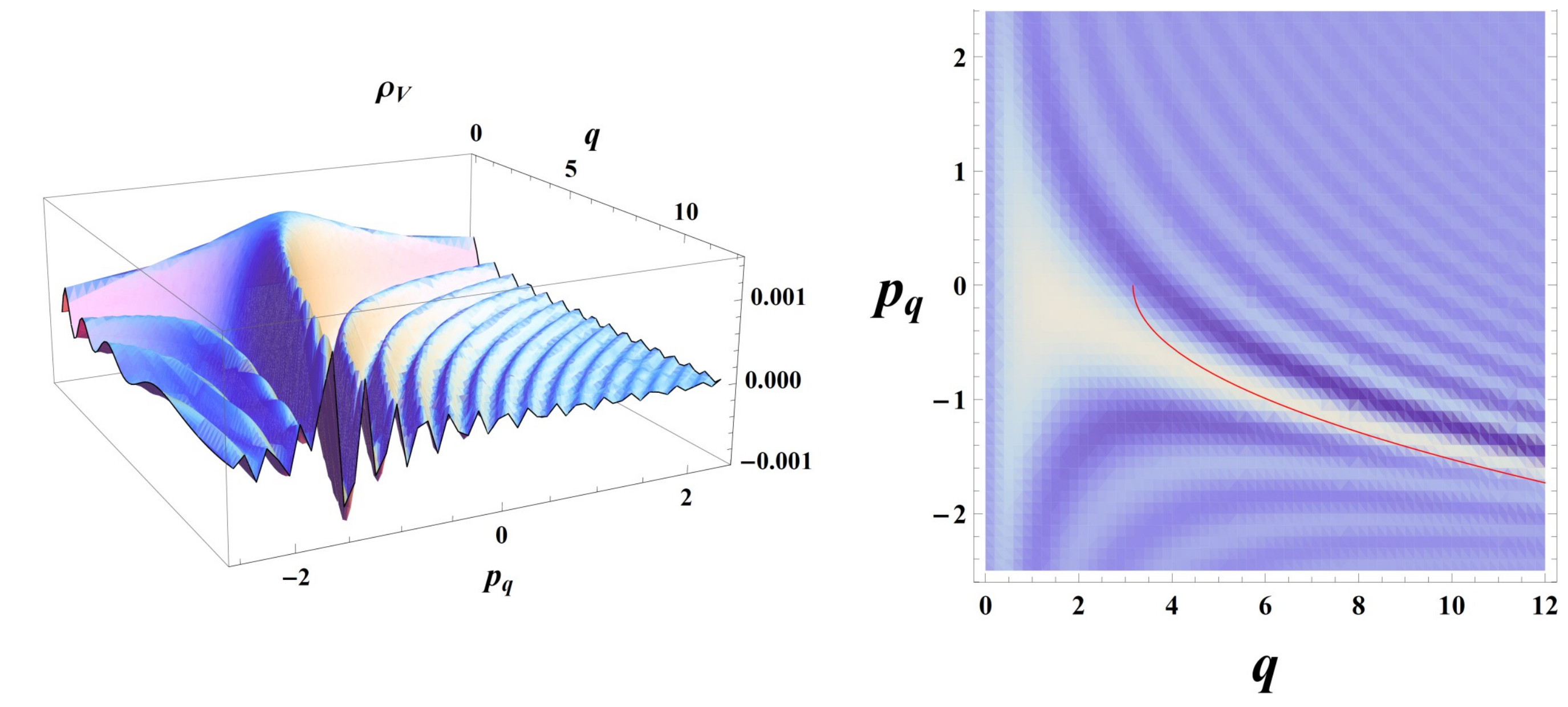}
\caption[Short caption for figure 1]{\label{fig:v468} {\scriptsize The Wigner function and its density plot for the Vilenkin boundary condition in a scenario where there is a potential barrier without embryonic era ($\hbar=1$). It can be observed a clear maximum and less oscillations compared with the Hartle-Hawking and Linde cases. The density projection illustrates that the classical trajectory is on the maxima of the Wigner function and has only one branch.}}
\end{center}
\end{figure}

For the Vilenkin boundary condition it can be appreciated only one branch
in its Wigner function (see Fig. \ref{fig:v468}) corresponding to an expanding Universe and its
highest peaks present a higher amplitude but with less oscillations than for the embryonic epoch
case. In Fig. \ref{fig:v468} it can be observed that the highest
oscillations of the Wigner function lies on the classical
trajectory. This fact can be explained again in terms of the decoherence of
the Vilenkin Wigner function as it happened in the previous case.

It is important to remark now that for the Linde and Vilenkin Wigner functions the highest peak
is near $q=0$ but this is not the case for general relativity. This difference can be explained
because the potential of the Wheeler-DeWitt equation tends to minus infinity when $q$ tends
to zero unlike general relativity which tends to a finite constant. For the Hartle-Hawking Wigner function
the peak of greater amplitude has a steeper slope that its counterpart in general relativity.


\noindent $(c)$
{\it No tunneling with big bang (Blue Region)}

\begin{figure}
\begin{center}
\includegraphics[scale=0.53]{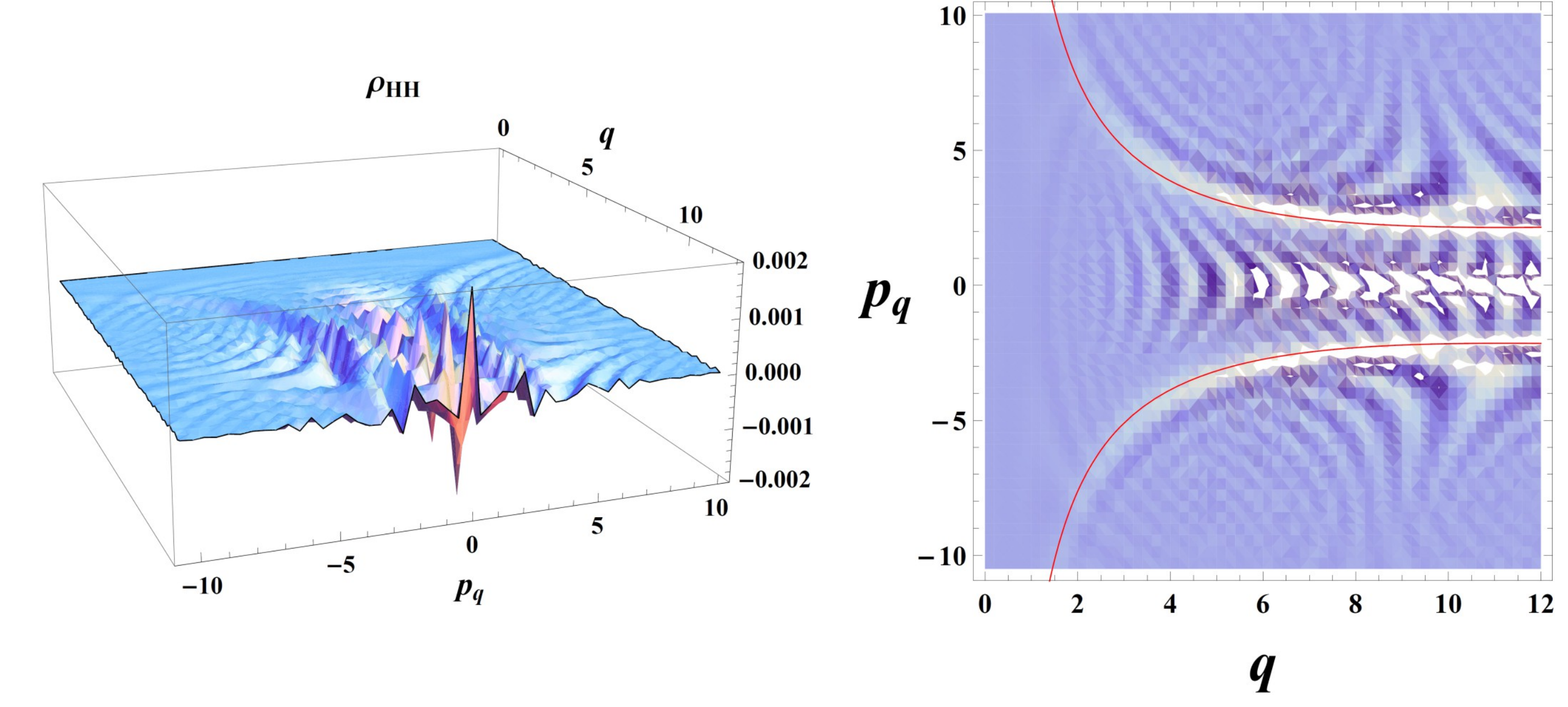}
\caption[Short caption for figure 1]{\label{fig:hh234} {\scriptsize The Wigner function and its density plot for the Hartle-Hawking case where a potential barrier is not present and the origin of the Universe is accessible ($\hbar=1$). In the figure it can be appreciated many oscillations due to the interference between wave functions of expanding and contracting universes. In its density projection it can be observed that the classical trajectories coincide with some of the highest peaks of the Wigner function.}}
\end{center}
\end{figure}

\begin{figure}
\begin{center}
\includegraphics[scale=0.53]{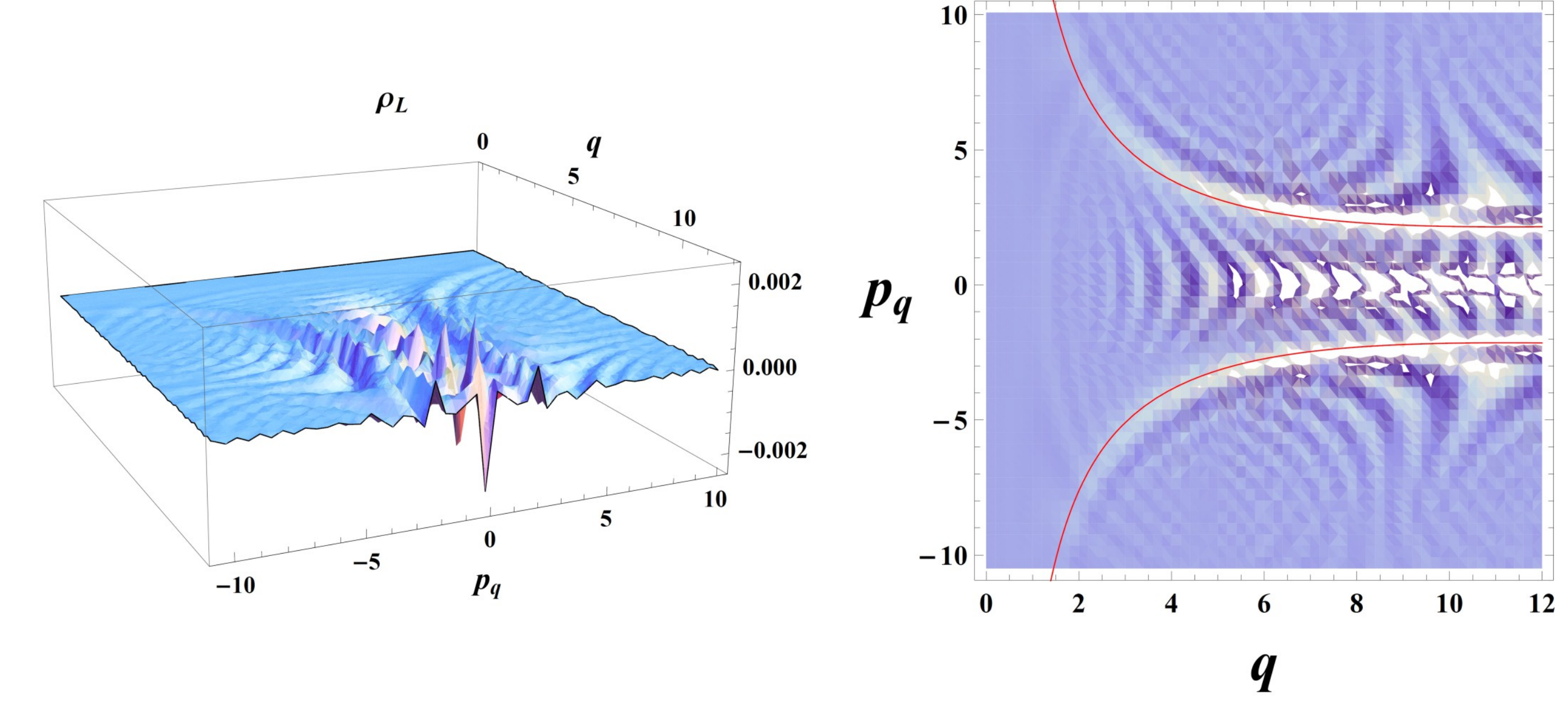}
\caption[Short caption for figure 1]{\label{fig:l234} {\scriptsize The Wigner function and its density plot for the Linde case where a potential barrier is not present and the origin of the Universe is accessible ($\hbar=1$). The figure shows a very similar behavior to the Hartle-Hawking boundary condition. In the density projection it can be appreciated that the highest peaks of the Wigner function are outside the classical trajectories.}}
\end{center}
\end{figure}

The last example corresponds to the following selection of values $g_2$ = $g_3$ = 0 and $g_4$ = $g_5$ = $g_6$ = 0.5 therefore $g_r$ = 0 and $g_s$ = 234
which give a point in the blue region of Fig. \ref{fig:region}.
In this case a potential barrier is not present and therefore a
tunneling process for the creation of the Universe is not possible.
This case corresponds to a contraction or an expansion process of the
Universe where the zero value of the scale factor is accessible. The expansion of the Universe
corresponds to the lower open curve in the density plots of the Wigner functions which have
negative values of the momenta while the upper open curve represents the contraction of the Universe.
A very similar behavior can be appreciated for the Hartle-Hawking and Linde
boundary conditions (see Fig. \ref{fig:hh234} and Fig. \ref{fig:l234}).
Both the Hartle-Hawking and the Linde Wigner functions posses many fluctuations
and some of its highest peaks are on the classical trajectories as can be appreciated
respectively also in Fig. \ref{fig:hh234} and Fig. \ref{fig:l234}.

\begin{figure}
\begin{center}
\includegraphics[scale=0.53]{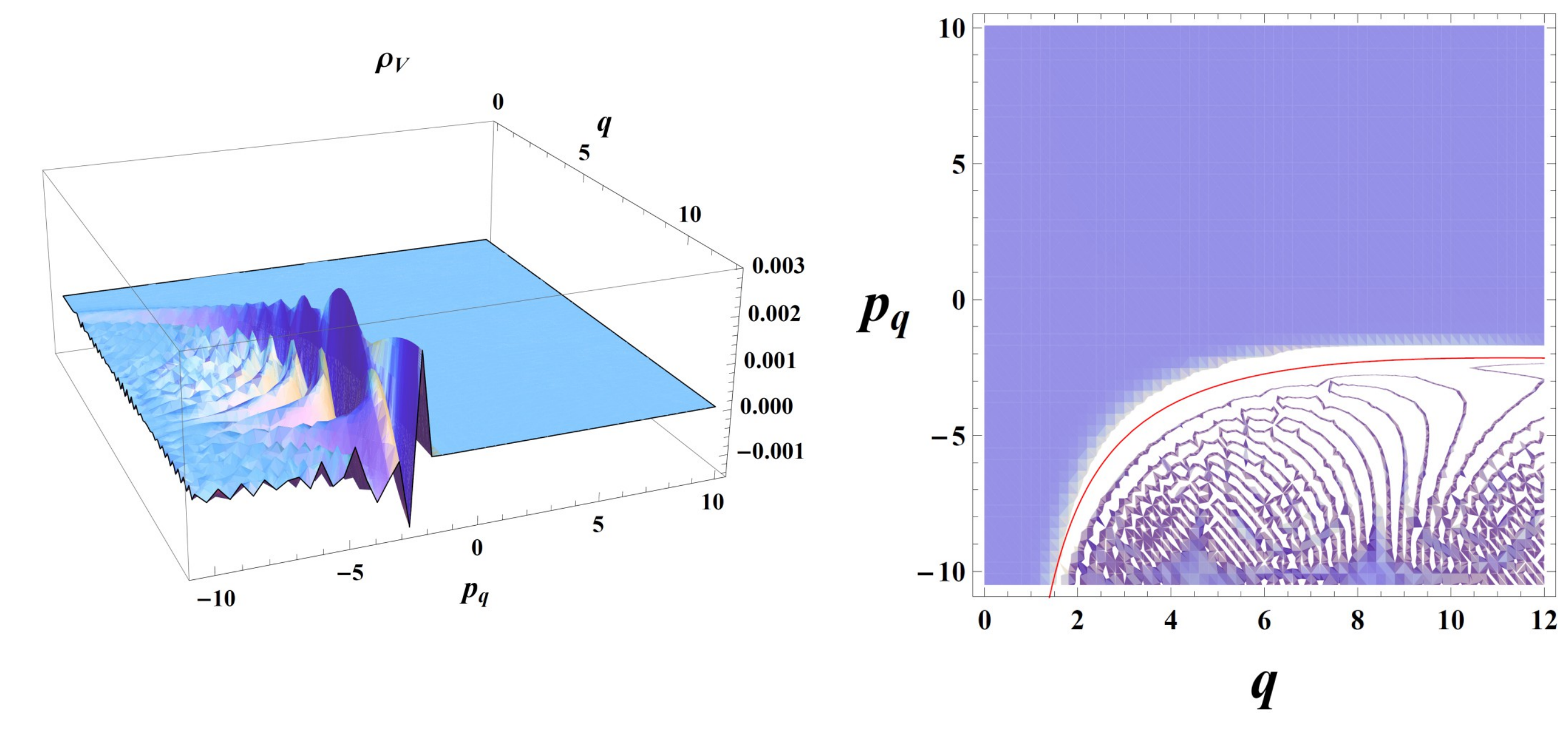}
\caption[Short caption for figure 1]{\label{fig:v234} {\scriptsize The Wigner function and its density plot for the Vilenkin case where a potential barrier is not present and the origin of the Universe is accessible ($\hbar=1$). It is observed a clear maximum and less oscillations compared with the Hartle-Hawking and Linde cases. The density projection illustrates that the classical trajectory is on the maxima of the Wigner function and has only one branch.}}
\end{center}
\end{figure}

\begin{figure}
\begin{center}
\includegraphics[scale=0.70]{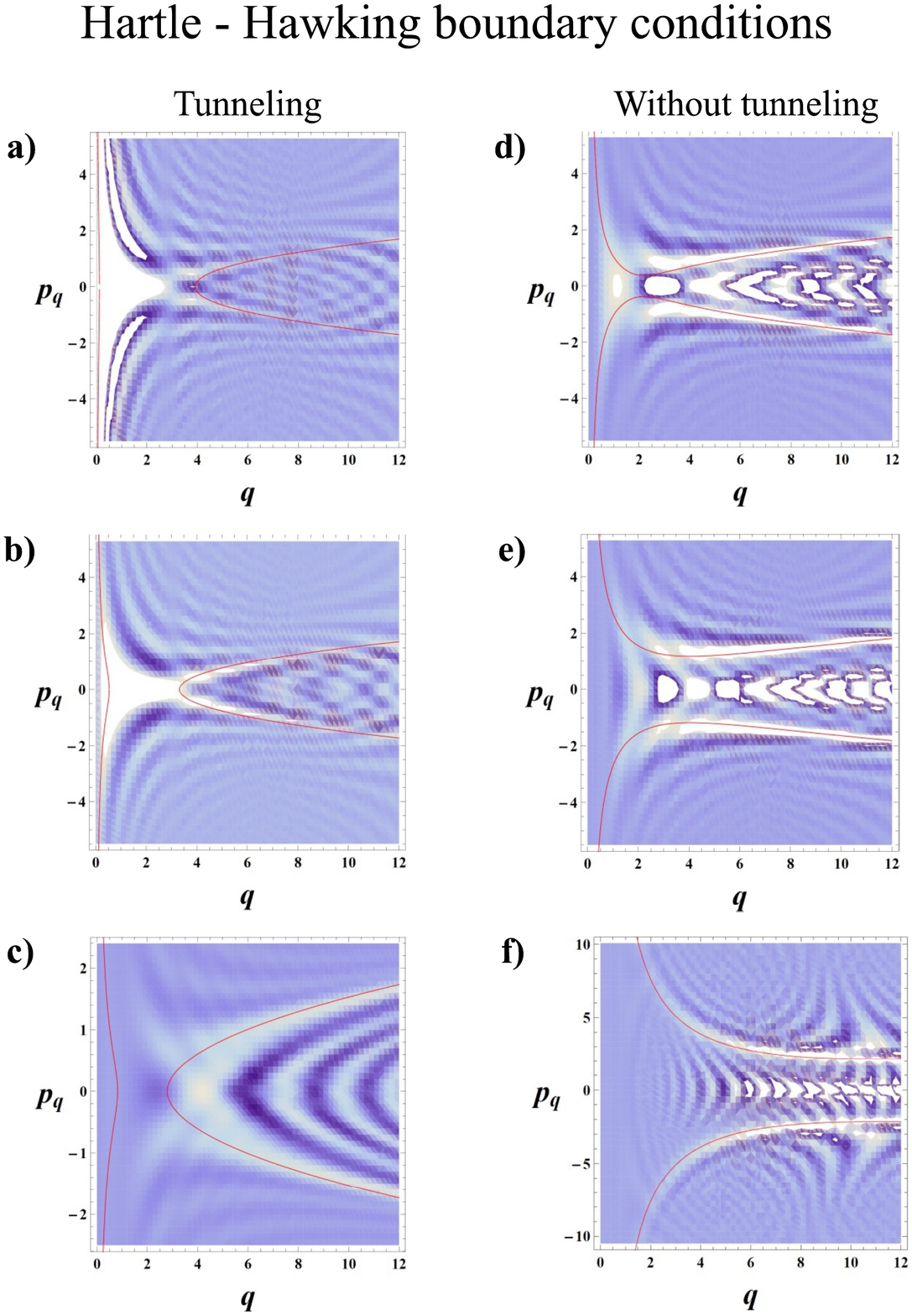}
\caption[Short caption for figure 1]{\label{fig:evoHH} {\scriptsize The Wigner function density plot for the Hartle-Hawking boundary condition ($\hbar=1$). On the left side three cases corresponding to a tunneling barrier are shown for the following values of parameters a) $g_r=-1.22$, $g_s = 0.15$, b) $g_r= -0.5$, $g_s = 0.5$, c) $g_r= 0.024$, $g_s = 0.468$. While on the right side three cases without tunneling barrier are displayed for the parameter values of d) $g_r= 0.21$, $g_s = 1.5$, e) $g_r=3$, $g_s = 5$ and f) $g_r= 0$, $g_s = 234$.}}
\end{center}
\end{figure}

\begin{figure}
\begin{center}
\includegraphics[scale=0.85]{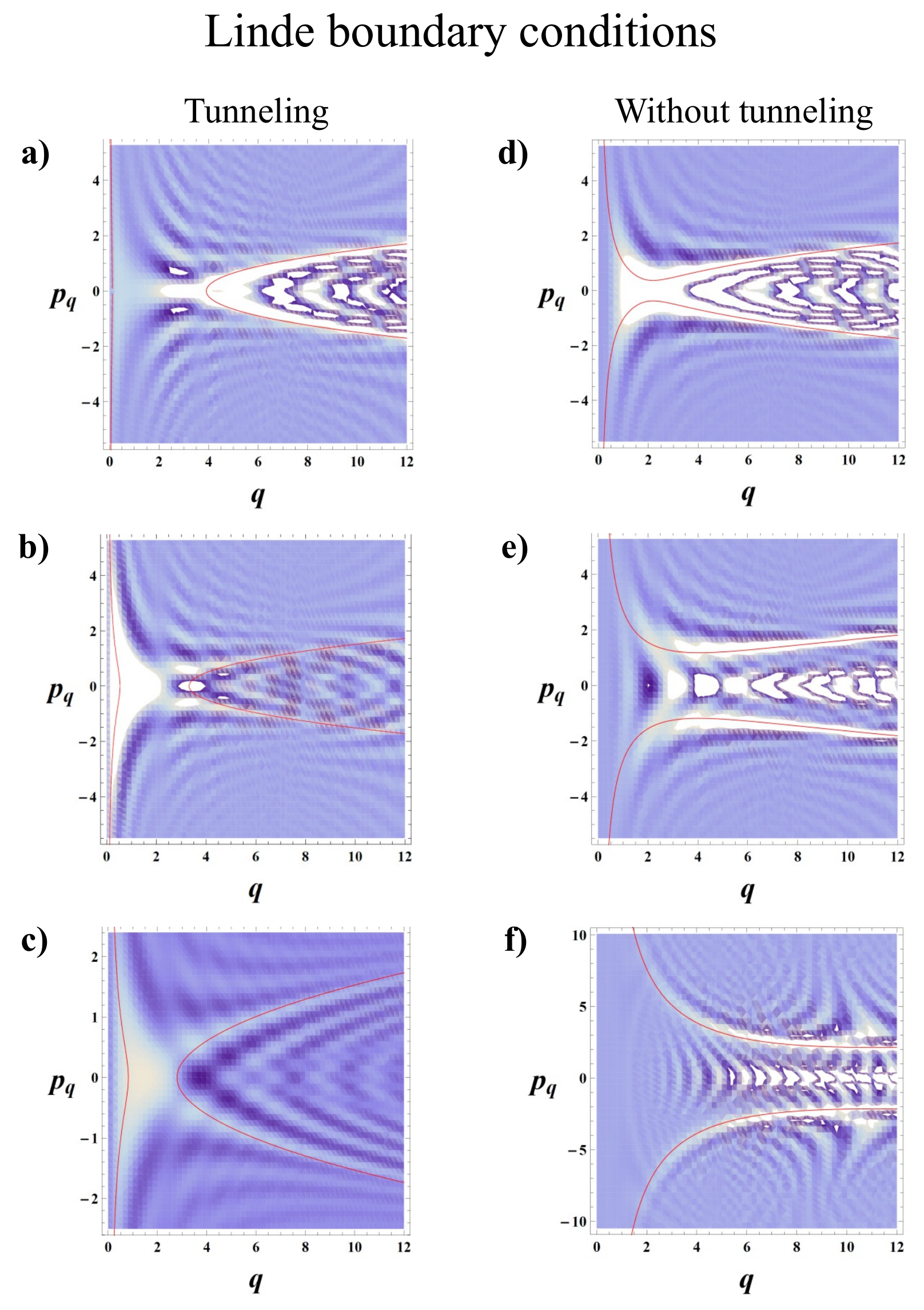}
\caption[Short caption for figure 1]{\label{fig:evoL} {\scriptsize The Wigner function density plot for the Linde boundary condition ($\hbar=1$). On the left side three cases corresponding to a tunneling barrier are shown for the following values of parameters a) $g_r=-1.22$, $g_s = 0.15$, b) $g_r= -0.5$, $g_s = 0.5$, c) $g_r= 0.024$, $g_s = 0.468$. The right side present three cases without tunneling barrier for the parameter values of d) $g_r= 0.21$, $g_s = 1.5$, e) $g_r=3$, $g_s = 5$ and f) $g_r= 0$, $g_s = 234$.}}
\end{center}
\end{figure}

\begin{figure}
\begin{center}
\includegraphics[scale=0.85]{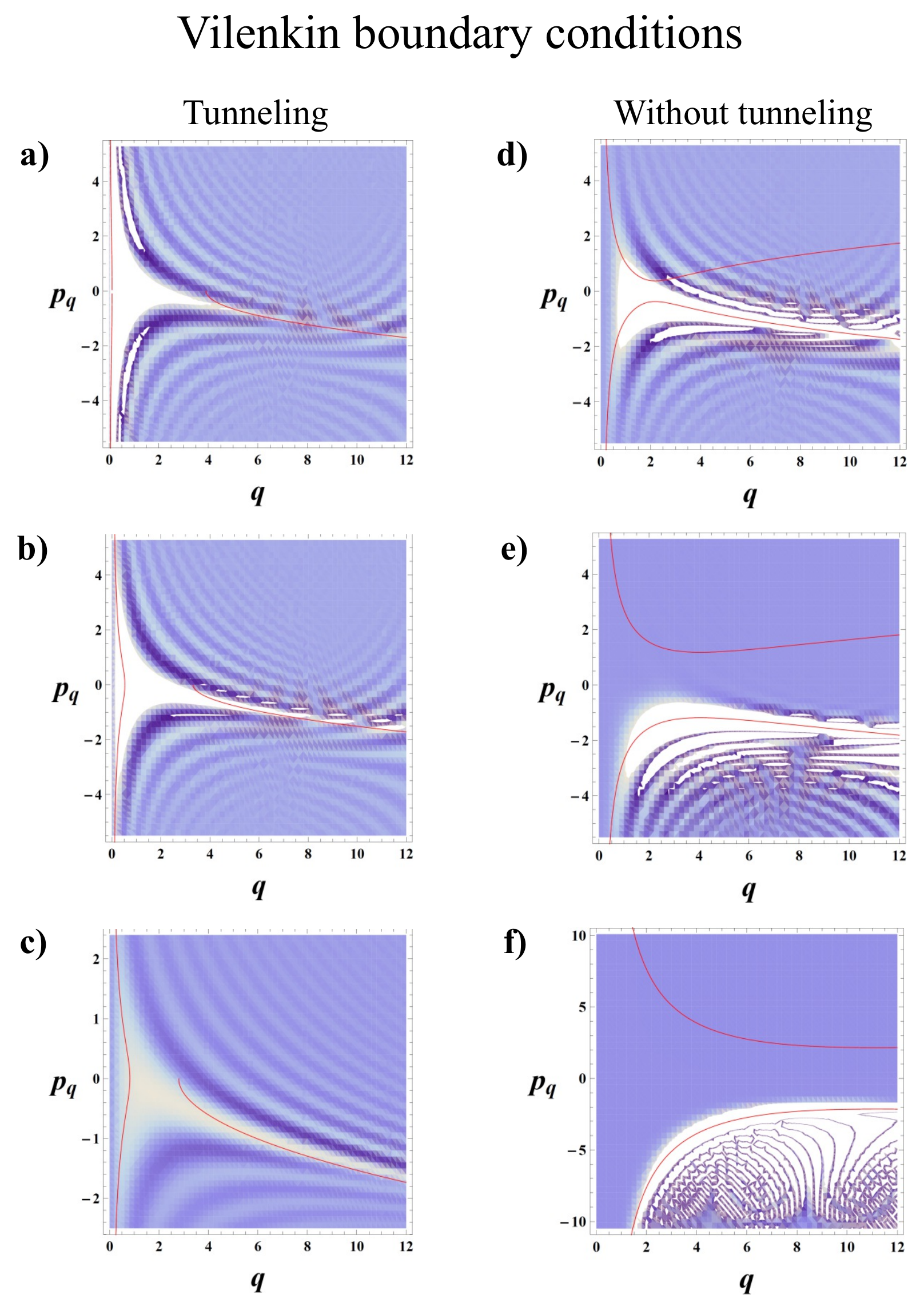}
\caption[Short caption for figure 1]{\label{fig:evoV} {\scriptsize The Wigner function density plot for the Vilenkin boundary condition ($\hbar=1$). On the left side three cases corresponding to a tunneling barrier are shown for the following values of parameters a) $g_r=-1.22$, $g_s = 0.15$, b) $g_r= -0.5$, $g_s = 0.5$, c) $g_r= 0.024$, $g_s = 0.468$. While on the right side three cases without tunneling barrier are displayed for the parameter values of d) $g_r= 0.21$, $g_s = 1.5$, e) $g_r=3$, $g_s = 5$ and f) $g_r= 0$, $g_s = 234$.}}
\end{center}
\end{figure}

\begin{figure}
\begin{center}
\includegraphics[scale=0.83]{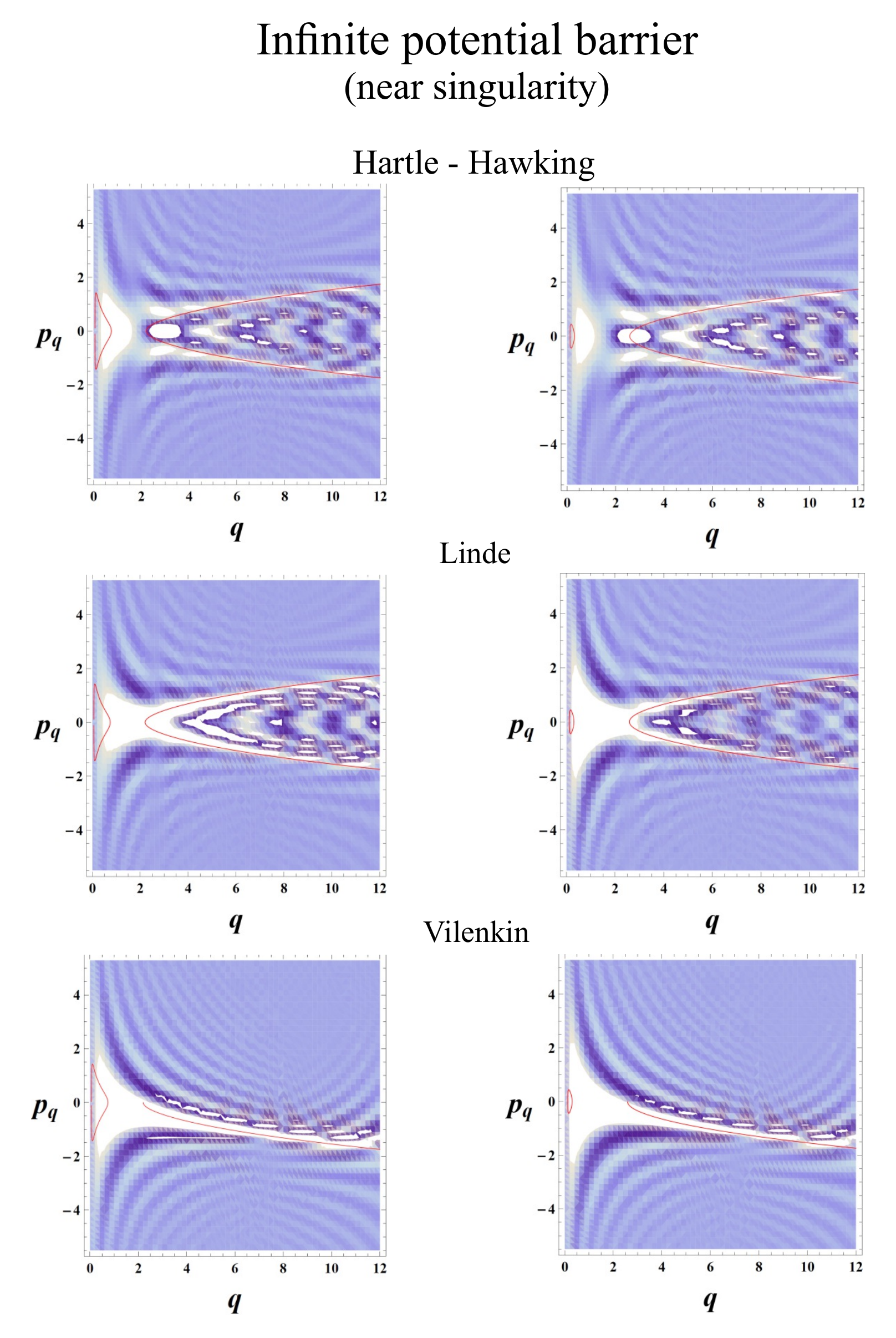}
\caption[Short caption for figure 1]{\label{fig:evoNBB} {\scriptsize The Wigner function density plot for the Hartle-Hawking, Linde and Vilenkin boundary conditions when there is an infinite potential barrier near the singularity ($\hbar=1$). The left column corresponds to the parameter values of $g_r= 0.6$ and $g_s = -0.03$. On the right column the parameters have the values of $g_r= 0.37$ and $g_s = -0.03$.}}
\end{center}
\end{figure}

Finally, for the Vilenkin boundary condition there is only one
branch corresponding to an expanding Universe where the highest
peaks of the Wigner function are located over the classical
trajectory. Furthermore, there are not oscillations above the
classical trajectory which is expected because this region
corresponds to a contracting Universe, see Fig. \ref{fig:v234}.
It is quite interesting to note that the region without any oscillations is a
consequence of the absence of a potential barrier which produces in the other
cases some interference effects in the corresponding region of the expansion
of the Universe (compare with Figs. \ref{fig:vnbb} and \ref{fig:v468}). Since
the interference between expanding and collapsing universes for this boundary
condition does not exist the decoherence of the Vilenkin Wigner functions seems to
be more convenient to achieve than the two other boundary conditions.

In order to illustrate the differences in the behavior of the Wigner function when the parameters $g_{r}$ and $g_{s}$ change we select six different values of them. These parameters are choose with the following purpose. They start with values which produce a high potential barrier which decreases until reaching a very small barrier. Subsequently, we consider values where the potential barrier is no longer present, starting with a maximum potential value very close to zero and then cases where such maximum value decreases.
In figures \ref{fig:evoHH}, \ref{fig:evoL} and \ref{fig:evoV} we study the three different boundary conditions and show the changes in the density plots for the Wigner function corresponding to a high tunneling barrier in (a), a medium tunneling barrier in (b), a small tunneling barrier in (c). Then, without tunneling barrier but with its maximum potential value near to zero in (d), without tunneling barrier but its maximum potential value more negative in (e) and finally in (f) the negative maximum potential is more distant from its zero value than the other two former cases. The values of the parameters employed in these three figures are given as follows. For (a) $g_2$ = $g_4$ = $g_5$ = 0, $g_3$ = -0.2033 and $g_6$ = 0.0041666 then $g_r$ = -1.22 and $g_s$ = 0.15. For (b) $g_2$ = $g_4$ = $g_5$ = 0, $g_3$ = -0.08333 and $g_6$ = 0.013889 generating the values $g_r$ = -0.5 and $g_s$ = 0.5. For (c) $g_2$ = $g_3$ = $g_4$ = $g_5$ = $g_6$ = 0.001 that produce $g_r$ = 0.024 and $g_s$ = 0.468. For (d) $g_2$ = $g_4$ = $g_5$ = 0, $g_3$ = 0.035 and $g_6$ = 0.041667 resulting in $g_r$ = 0.21 and $g_s$ = 1.5. For (e) $g_2$ = $g_4$ = $g_5$ = 0, $g_3$ = 0.5 and $g_6$ = 0.138889 then $g_r$ = 3 and $g_s$ = 5. And for (f) $g_2$ = $g_3$ = 0 and $g_4$ = $g_5$ = $g_6$ = 0.5 so that $g_r$ = 0 and $g_s$ = 234.

In the case of the Hartle-Hawking boundary condition with tunneling we can observe that the higher value of the Wigner function is located between the close and open classical trajectories but when the potential barrier value decreases the higher value of the Wigner function shift to the right until is situated between the two branches of the open trajectories. For the Linde boundary condition the opposite situation happens. At the beginning the higher value of the Wigner function is inside the open classical trajectories corresponding to the tunneling process however when the potential barrier reduces its value the maximum of the Wigner function
shift to the left and it is placed between the close and open classical trajectories. On the right side, when there is no potential barrier and the potential maximum value is very close to zero, the Wigner function highest peak for the Hartle-Hawking boundary is located in a small region between the two classical trajectories but for the Linde case it is placed in a broader region. When the maximum value of the potential is more negative, the highest peak of the Wigner function is nearer the singularity for the Hartle-Hawking boundary condition than for the Linde option. In this case, the Linde Wigner function presents more oscillations than the Hartle-Hawking function. The Hartle-Hawking and Linde cases are very similar when the maximum value of the potential is too negative.

For the Vilenkin boundary condition we can appreciate that the highest value of the Wigner functions is near the initial singularity when the potential barrier is high and moves away of the singularity when the potential barrier decreases its value. In the case when there is not potential barrier and its maximum value is near to zero, the highest peak of the Wigner function is located in a broad region near the singularity but when the maximum of the potential value decreases the higher peaks are located on the classical trajectory corresponding to an expanding Universe.

In figure \ref{fig:evoNBB} on the left side, we present the density plot for the Wigner functions when there is a infinite potential barrier near the singularity with a deep well (embryonic epoch) and a small tunneling barrier while on the right side the density plots of the Wigner functions correspond to a slight well and a high tunneling barrier. The values of the parameters employed are as follows. On the left side $g_2$ = $g_4$ = $g_5$ = 0, $g_3$ = 0.1 and $g_6$ = -0.0008333 producing the values $g_r$ = 0.6 and $g_s$ = -0.03. While on the right side $g_2$ = $g_4$ = $g_5$ = 0, $g_3$ = 0.0616667 and $g_6$ = -0.0008333 producing the values $g_r$ = 0.37 and $g_s$ = -0.03. In these graphics one can appreciate that the highest values of the Wigner function are located in a broader region between the two classical trajectories for the slight well and high tunneling barrier than for deep well and small tunneling barrier for the cases of Linde and Vilenkin boundaries conditions. For the Hartle-Hawking case the opposite situation happens. Finally, the higher peaks of the Wigner functions shift to the initial singularity when there is a slight well and a high tunneling barrier than when there is a deep well and a small tunneling barrier.

\section{Final Remarks}

The Ho$\check{\rm r}$ava-Lifshitz gravity has many appealing
features like being a renormalizable theory. In this way, the
Ho$\check{\rm r}$ava-Lifshitz quantum cosmology has richer
characteristics than the FLRW Einstein quantum cosmological model.
For example, the two extra terms in the quantum potential give rise
to the existence of a possible embryonic epoch where the Universe
can exist classically oscillating between two small values of the
scale factor. This situation does not take place in usual general relativity.
It is very interesting to note that for this case the singularity is not accessible
in the classical regime due to an infinite potential barrier (see purple curve
in figure 1).

In order to analyze the quantum aspects of the early Universe we
study their Wigner functions since they provide a different perspective of
the system in the phase space.
In this article we studied the quantum behavior of the
FLRW model in HL type gravity by means of the Wigner functions
satisfying three different boundary conditions implemented at large values of
the scale factor. These correspond to the Hartle-Hawking, Linde and
Vilenkin proposals. For the Hartle-Hawking and Linde cases the
quantum description of the Universe has a contracting an expanding
components of the Universe that give rise to interference patterns
which presents many oscillations near the classical trajectory.
While the Vilenkin proposal corresponds only to an expanding
Universe and it can be associated to a tunneling process.

Using the four terms in the quantum potential (see Eq. (\ref{WDWPOTENTIAL}))
we have investigated three different scenarios corresponding to three particular regions
of Fig. \ref{fig:region}.

The first one $(a)$ describes the situation where a tunneling process and an embryonic epoch are possible ({\it Purple Region}).

In the embryonic era the Universe can exist classically oscillating between two small values of the
scale factor. In this scenario the Universe can nucleate to a finite
size by means of a tunneling process and after that it can be
expanded along the usual lines of the inflationary model of our
Universe. This is an interesting case for the early evolution of the Universe which it does not
appear in usual general relativity quantum cosmology. Moreover it can be observed a
highest peak of the Wigner function near the zero value of the scale factor which is consistent
with the oscillation of the Universe between two different no null
values of the scale factor. Besides, it is possible to note that
there exist a tunneling process. For the
Hartle-Hawking and Linde boundary conditions there are two branches
of the classical trajectory corresponding to a contracting and
expanding universes, and the highest oscillations are on these
curves. The Hartle-Hawking case presents more fluctuations of the Wigner function between the two classical regions near $P_q=0$
than the Linde case. For the Vilenkin boundary condition there is only one classical trajectory corresponding to an
expanding Universe (negative values of the momenta) which it is in agreement with tunneling boundary condition.
The higher peaks of the Wigner functions are closer to the initial singularity for the case of a slight well and a high tunneling barrier than for a deep well and a small tunneling barrier. In addition, some of the higher fluctuations of the Wigner function are on the classical trajectory that it can be explained
because there are not interference terms between expanding and contracting universe
and the decoherence is easier to achieve than for the two other boundary conditions.

In the case (b) of tunneling without embryonic epoch ({\it Green Region}) we observe that the initial singularity is
classically and quantum accessible. The Universe can exist classically with non zero value of the scale factor before
tunneling. This scenario is not present in usual quantum cosmology and constitutes a new feature of HL quantum cosmology. Once the tunneling process takes place the Universe can
evolve according to the inflationary paradigm and expand along the established by the standard cosmological model. For the Hartle-Hawking boundary condition with a high tunneling barrier the highest value of the Wigner function is located between the close and open classical trajectories but when the potential barrier maximum value decreases the highest value of the Wigner function moves to the right until is placed between the two branches of the open classical trajectories for a small barrier. For the Linde case the opposite situation appears: the highest value of the Wigner function is inside the open classical trajectories for a high tunneling barrier but when the tunneling potential reduces its value the maximum of the Wigner function moves to the left and it is placed between the close and open classical trajectories. This difference can be understood because the interference terms have opposite signs for these boundary conditions. For both cases the next higher peaks are on the classical trajectory but the amplitudes of the fluctuations are
bigger and present more oscillations for Linde case than the Hartle-Hawking boundary condition. The Wigner function for Vilenkin boundary condition presents only one branch corresponding to an expanding Universe. Furthermore, it can be appreciated that the Wigner function has a higher amplitude and less oscillations when it is compared with the general relativity case (see \cite{Cordero:2011xa}). Another important difference with general relativity quantum cosmology is that the highest peak of the Wigner function is near $q=0$, and it can be explained because for the HL quantum cosmology the Wheeler-DeWitt potential diverges to minus infinity at $q=0$.

The (c) case of no tunneling with big bang ({\it Blue Region}) corresponds to an scenario which is similar to a
dispersion process where a potential barrier is not present. This situation produces an expansion or
contraction of the Universe where the initial singularity is accessible at classical and quantum levels.
In this case, for the three boundary conditions studied there is not a highest peak of the Wigner
function near the zero value of the scale factor. When the maximum value of potential is very close to zero, the Wigner function highest peak for the Hartle-Hawking boundary is placed in a small region between the two classical trajectories but for the Linde case it is located in a broader region. For a more negative value of the potential maximum, the highest peak of the Wigner function is nearer the singularity for the Hartle-Hawking boundary condition than for the Linde case. It can be appreciated that the Hartle-Hawking and Linde Wigner functions present a very similar behavior when the potential maximum value is too negative. Besides, some of the higher peaks of the Wigner function are on the classical trajectory.

In all the three cases the size of the fluctuations are of the same order but for the Vilenkin boundary condition there are not fluctuations in the region where the momenta are positive. It is very interesting to note that the region with no fluctuations of the Wigner function is a consequence of the absence of a potential barrier which produces, in the other cases analyzed before, interference effects in the region that describes the expansion of the Universe. For the Vilenkin boundary condition there is not contracting Universe and the decoherence seems to be obtained more easily.

We want to stress that it is relevant to investigate the role and effects of
the different boundary conditions on the physical behavior of the
Universe. Among the physical effects on the behavior of the
Universe are the presence of inhomogeneities in ground states that
could fit cosmic microwave background radiation data
\cite{Page:2006mz} and the existence of possible initial conditions
for inflation \cite{Calcagni:2015vja}. However, the issue of which
one is the right boundary condition for the Universe is open to
debate from long time ago \cite{debate}. For example, in some papers
it was argued that the Hartle-Hawking boundary condition predicts
small amount of inflation  and the tunneling boundary condition
gives a large amount \cite{Calcagni:2015vja,Vilenkin:1987kf}. It was
claimed too that the Hartle-Hawking and tunneling boundary condition
have physical problems \cite{Page:2006mz,
Feldbrugge:2018gin,Feldbrugge:2017fcc}. In fact, this debate
continues in very recent papers
\cite{Vilenkin:2018dch,Vilenkin:2018oja} where it is claimed that
the possible runaway instabilities and the strong coupling problem
from the tunneling boundary condition are under control. The link
between the quantum description and the large scale properties of
the Universe is dependent of the assumptions about cosmological
boundary characteristics and the initial conditions for inflation
among others. However, the non existence of a complete definition
for the specific properties of initial quantum states
\cite{DiTucci:2019xcr} produces difficulties in order to achieve a
detailed modeling of the primordial conditions of the Universe. The
former results restrict the possibility to study observational
implications. Although the initial conditions for inflation might
not be possible to obtain from quantum cosmology it is very
important to analyze the different scenarios from the boundary
conditions in order to achieve a possible complete description for
the origin of the Universe.

It is important to mention that in addition to the theory of
Ho$\check{\rm r}$ava-Lifshitz there are other alternative proposals
of gravity that include modifications in the ultraviolet regime such
that allow to improve its quantum behavior. One of them is the
so-called Gravity's Rainbow which introduce changes directly to the
metric instead of modifying the action as in the Ho$\check{\rm r}$ava-Lifshitz
approach. This modified metric presents a different
treatment between space and time in the UV regime as it happens with
Ho$\check{\rm r}$ava-Lifshitz. However, at low energies it
allows to recover the usual general relativity. The construction of
this proposal can be consulted in \cite{Magueijo:2002xx}. Later, an
interesting and detailed analysis of the relationship between
Gravity's Rainbow and Ho$\check{\rm r}$ava-Lifshitz was carried out
in \cite{Garattini:2014rwa}. In that work it is found a
correspondence between Gravity's Rainbow and the theory of
Ho$\check{\rm r}$ava-Lifshitz. Such relationship was obtained through the
Wheeler-DeWitt equations corresponding to both gravitational
proposals and it was done for the case of FLRW and for geometries
with spherical symmetries. This way of using the Wheeler-DeWitt equation
can be employed to study the relation between other alternative
gravitational theories and Ho$\check{\rm r}$ava-Lifshitz gravity.  Finally,
given the relationship found in \cite{Garattini:2014rwa}, it will be
also interesting to study the Gravity's Rainbow proposal through
Wigner functions and compare those results with the ones obtained in
this work which could give a deeper insight between both gravity
theories.

The description of this system through the Wigner function even
numerically allows to obtain novel results as the possibility to
analyse an embryonic epoch of the Universe from a different perspective.
It would be interesting to apply this approach to explore other cosmological models in the
Ho$\check{\rm r}$ava-Lifshitz type gravity as well as in other
models of modified gravity.

\vskip 1truecm

\centerline{\bf Acknowledgments}

\vskip 1truecm

The work of R. C., H. G.-C. and F. J. T. was partially supported by SNI-M\'exico, CONACyT research grant: 128761. In addition R. C. and F. J. T. were partially supported by COFAA-IPN and by SIP-IPN grants 20171168, 20171100, 20180735, 20180741, 20194924 and 20195330. We are indebted to H\'ector Uriarte for all his help in the elaboration of the figures presented in the paper.


\end{document}